\documentclass[aps,prd,twocolumn,showpacs]{revtex4}
\usepackage{bm,amsfonts,amssymb,amsmath}
\begin{document}
    \title{Universal Features of Dimensional Reduction Schemes \protect\\
    from General Covariance Breaking}
\author{Paolo~\surname{Maraner$^1$} and Jiannis~K.~\surname{Pachos$^{2,3}$}}
\affiliation{$^1$School~of~Economics~and~Management,
Free~University~of~Bozen-Bolzano,~via~Sernesi~1,~39100~Bolzano,~Italy\\
$^2$School of Physics and Astronomy, University of Leeds, Leeds LS2 9JT,
UK\\
$^3$Kavli Institute for Theoretical Physics, University of California, Santa
Barbara, CA 93106, USA}
\date{\today}

\begin{abstract}
Many features of dimensional reduction schemes are determined by the
breaking of higher dimensional general covariance associated with the
selection of a particular subset of coordinates. By investigating residual
covariance we introduce lower dimensional tensors, that successfully
generalize to one side Kaluza-Klein gauge fields and to the other side
extrinsic curvature and torsion of embedded spaces, thus fully
characterizing the geometry of dimensional reduction. We obtain general
formulas for the reduction of the main tensors and operators of Riemannian
geometry. In particular, we provide what is probably the maximal possible
generalization of Gauss, Codazzi and Ricci equations and various other
standard formulas in Kaluza-Klein and embedded spacetimes theories. After
general covariance breaking, part of the residual covariance is perceived by
effective lower dimensional observers as an infinite dimensional gauge
group. This reduces to finite dimensions in Kaluza-Klein and other few
remarkable backgrounds, all characterized by the vanishing of appropriate
lower dimensional tensors.
\end{abstract}

\pacs{02.40.-k, 04.50.+h} \maketitle

\section{Introduction}

 In many different situations, ranging from low to high energy
physics, we are interested in --or have access to-- only a part of
the coordinates describing a given physical system. The problem is
finding the effective dynamics that drive the interesting
variables by reducing the uninteresting ones or, vice versa, given
the effective dynamics of accessible variables, introducing extra
coordinates that simplify the overall dynamical picture.
Dimensional reduction may be induced by a number of very different
mechanisms which in general leave a track in the lower dimensional
dynamics. Typical examples are Kaluza-Klein and brane-world
reduction in high energy physics, quantum dots/lines/surfaces in
semiconductor physics, magnetic confinement in plasma physics and
so on. However, there are features that only depend on the
selection of the `interesting' coordinates and not on the specific
mechanism under consideration. In this paper we focus on these
universal features that depend on the selection of a subset of
coordinates and not on specific reduction schemes. It should be
stressed, that even if in certain cases --like brane-worlds and
quantum lines/surfaces-- the effective lower dimensional
configuration space can be identified with a regularly embedded
metric submanifold, this is not the general situation. A classical
example is provided by Kaluza-Klein theories where the physical
spacetime is obtained by identifying higher dimensional points
connected by a special class of diffeomorphisms that will
eventually be identified with gauge transformations. The resulting
quotient space can not be given the structure of metric
submanifold. The classical theory of embeddings
\cite{submanifolds,embST} is not enough for describing the general
situation. In this paper we further investigate the geometry of
coordinate separation with emphasis on residual general covariance
and provide a unifying framework that successfully generalizes the
theory of metric submanifolds.

The paper is organized as follows. In Section \ref{sec1} we find
that dimensional reduction is completely characterized by lower
dimensional tensors, generalizing, on the one hand, Kaluza-Klein
gauge fields~\cite{KK,nonAbKK,KKrev} and, on the other, extrinsic
curvature and torsion --i.e.\ second and normal fundamental
forms-- of embedded spaces \cite{submanifolds,embST}. In terms of
these we obtain in Section \ref{sec2} general reduction formulas
for the Riemann tensor, Ricci tensor, scalar curvature, geodesics
equations, Laplace and Dirac operators, providing what is probably
the maximal possible generalization of Gauss, Codazzi,  Ricci
equations \cite{GCR} and various other standard identities in
embeddings and Kaluza-Klein theories. These equations also
represent the natural starting point to investigate higher
dimensional unification scenarios in which physics is allowed to
fully depend on all the introduced coordinates. In
Section~\ref{sec3} special attention is given to induced gauge
structures. We show how residual general covariance in the reduced
variables always emerges in the effective dynamics as gauge
covariance. The induced gauge group is in general infinite
dimensional and reduces to finite dimensions in Kaluza-Klein and a
few other remarkable backgrounds, all characterized by the
vanishing of appropriate lower dimensional tensors. Finally, in
Section~\ref{sec4} a discussion of the findings is presented and
concluding remarks are made.

For the shake of concreteness we tackle the problem from the viewpoint of
higher dimensional unification. We consider a higher dimensional (HD)
spacetime ${\mathbf M}_D$ parameterized by $D$ continuous coordinates
${\mathbf x}^I$, $I=0,1, ... ,D-1$, endowed with a pseudo-Riemannian metric
${\mathbf g}_{IJ}$. In addition to the coordinate system, we set up
reference frames at each spacetime point ${\mathbf r}_A^{\ I}({\mathbf x})$,
$A=0,1, ... ,D-1$, ${\mathbf r}_A^{\ I} {\mathbf r}_B^{\ J}{\mathbf
g}_{IJ}=\eta_{AB}$. Physical laws are assumed to be covariant under general
coordinates transformations and local redefinitions of reference frames
\cite{Weinberg72}
\begin{equation}
{\mathbf x}^I \rightarrow {\mathbf x}'^I({\mathbf x})\hskip0,5cm
{\mathbf r}_A^{\ I} \rightarrow {\mathbf\Lambda}_A^{\ B}({\mathbf
x}){\mathbf r}_B^{\ I} \label{genD}
\end{equation}
At low energies the spacetime ${\mathrm M}_d$ (e.g. $d=4$) is parameterized
by $d$ continuous coordinates $x^\mu$, $\mu=0,1,...,d-1$, and reference
frames are made up of $d$ reference vectors $r_\alpha^{\ \mu}$,
$\alpha=0,1,...,d-1$. Physical laws are covariant under (electroweak and
strong) gauge transformations, other than general coordinates
transformations $x^\mu \rightarrow x'^\mu(x)$ and local redefinitions of
reference frames $r_\alpha^{\ \mu}\rightarrow \Lambda_\alpha^{\ \beta}(x)
r_\beta^{\ \mu}$.
 The original motivation for considering higher dimensional
unification is the hope that HD covariance can account for lower
dimensional (LD) gauge symmetries other than LD spacetime
covariance.
 To make contact with LD physics, we split HD coordinates in two
groups ${\mathbf x}^I =(x^\mu,y^i)$ with $\mu=0,1,...,d-1$,
$i=1,2,...,c\equiv D-d$. We refer to $x^\mu$ and $y^i$ as {\em external} and
{\em internal} coordinates, respectively. Consequently, reference frames
split in four blocks ${\mathbf r}_\alpha^{\ \mu}\equiv r_\alpha^{\ \mu}$,
..., ${\mathbf r}_{a+d-1}^{\ i+d-1}\equiv \rho_a^{\ i}$ with
$\alpha=0,1,...,d-1$, $a=1,2,...,c$. As we are willing to make no a priori
hypothesis on specific reduction mechanisms, we proceed by noticing that the
minimal assumption that drives us to recover the desired LD spacetime
covariance is that the HD transformation group (\ref{genD}) is effectively
broken down to
\begin{equation}
\left\{
\begin{array}{l}
x^\mu\rightarrow x'^\mu(x)\\
y^i\rightarrow y'^i(x,y)
\end{array}
\right. \hskip0,5cm \left\{
\begin{array}{l}
r_\alpha^{\ \mu}\rightarrow \Lambda_\alpha^{\ \beta}(x)r_\beta^{\
\mu}\\\ \vdots\\ \rho_a^{\ i}\rightarrow \Lambda_a^{\
b}(x,y)\rho_b^{\ i}
\end{array}
\right.\label{STr}
\end{equation}
We take this as a characterization of dimensional reduction. In working out
the consequences that it implies, as a check of our results and to make
contact with the most important applications, we constantly specialize in
appropriate subsections to Kaluza-Klein theories \cite{KK,nonAbKK,KKrev} and
spacetimes embedded in a flat \footnote{The assumption of flatness is
clearly not necessary and is made because it is common in applications and
to keep our explanatory formulas as simple as possible.} higher dimensional
space~\cite{embST}. While in the former case the topology reduces to that of
a direct product and in the latter the system is localized on a submanifold,
in the general case the structure of the HD spacetime is more complex. In
correspondence to every choice of external coordinates $x^\mu$, the internal
coordinates $y^i$  span a $c$-dimensional {\em internal spacetime} ${\mathrm
M}_c^x$ regularly embedded in ${\mathbf M}_D$. Every internal spacetime
${\mathrm M}_c^x$ has to be identified with a point of the $d$-dimensional
{\em external spacetime} ${\mathrm M}_d$ and may posses a geometry --and
even a topology-- that vary from point to point. Strictly speaking,
${\mathrm M}_d$ can not be identified with the effective spacetime before
internal coordinates have been completely removed. In spite of this we will
talk about LD external metric, curvature or general tensors, with the bona
fide assumption that internal coordinate dependence will be eventually
removed from the effective theory. Clearly, any realistic reduction
mechanism will eventually involve such a removal. However we will not
address this issue in this paper.

%%%%%%%%%%%%%%%%%%%%%%%%%%%%%%%%%%%%%%%%%%%%%%%%%%%%%%%%%%%%%%%
%
\section{\label{sec1} The Geometry of \protect\\ Dimensional Reduction}
%
%%%%%%%%%%%%%%%%%%%%%%%%%%%%%%%%%%%%%%%%%%%%%%%%%%%%%%%%%%%%%%%%

The HD spacetime ${\mathbf M}_D$ is endowed with standard
pseudo-Riemannian geometry.
%%%%%%%%%%%%%%%%%%%%%%%%%%%%%%%%%%%%%%%%%%%%%%%%%%%%%%%%%%%%%%%%%%%%
\subsection{\label{sec1.1}Tensors}
%%%%%%%%%%%%%%%%%%%%%%%%%%%%%%%%%%%%%%%%%%%%%%%%%%%%%%%%%%%%%%%%%%%%
HD {\em tensors} ${\mathbf t}_{...I...}^{\ ...J...}$ transform
according to
\[
{\mathbf t}_{...I...}^{\ ...J...}\rightarrow ...\ {\mathbf
J}_{I}^{\ K} ...\ {\mathbf t}_{...K...}^{\ ...L...}\ ...\
{{\mathbf J}^{-1}}_{L}^{\ J}...
\]
with ${\mathbf J}_{I}^{\ J}=\frac{\partial{\mathbf
x}^J}{\partial{\mathbf x}'^I}$ the Jacobian matrix associated with
the transformation of HD coordinates.

LD {\em external tensors} $t_{...\mu...}^{\ ...\nu...}$  and  LD
{\em internal tensors} $t_{...i...}^{\ ...j...}$, respectively
carrying external and internal indices, transform according to
\begin{eqnarray*}
t_{...\mu ...}^{\ ...\nu...}&\rightarrow& ...\ J_{\mu}^{\
\kappa}...\
t_{...\kappa ...}^{\ ...\lambda...}\ ...\ {J^{-1}}_{\lambda}^{\ \nu}...\\
t_{...i...}^{\ ...j...}&\rightarrow& ...\ J_{i}^{\ k}...\
t_{...k...}^{\ ...l...}\ ...\ {J^{-1}}_{l}^{\ j}...
\end{eqnarray*}
with $J_\mu^{\ \nu}=\frac{\partial x^\nu}{\partial x'^\mu}$ and $J_i^{\
j}=\frac{\partial y^j}{\partial y'^i}$ the Jacobian matrices associated with
the transformations of $x^\mu$ and $y^i$ respectively. LD {\em hybrid
tensors} $t_{...\mu... i...}^{\ ...\nu... j...}$, carrying internal and
external indices that transform with $J_\mu^{\ \nu}$ and $J_i^{\ j}$,
respectively, will also be considered.

When HD covariance is broken from (\ref{genD}) to (\ref{STr}) ,
${\mathbf J}_{I}^{\ J}$ takes the block non-diagonal form
\begin{equation}
{\mathbf J}_{I}^{\ J}({\mathbf x}')= \left(
\begin{array}{cc}
J_\mu^{\ \nu}(x') & \frac{\partial y^j}{\partial x'^\mu}(x',y')
\\ 0
& J_i^{\ j}(x',y')\nonumber
\end{array}\right)
\end{equation}
The off-diagonal block  makes covariant external ${\mathbf t}_{...\mu...}$,
contravariant internal ${\mathbf t}^{...i...}$ and analogous hybrid
components ${\mathbf t}_{...\mu...}^{\ ...j ...}$ of HD tensors,  in
non-covariant LD objects. On the other hand, contravariant external
${\mathbf t}^{...\mu...}$, covariant internal ${\mathbf t}_{...i...}$  and
analogous hybrid components ${\mathbf t}_{...i...}^{\ ...\nu...}$ of HD
tensors, transform like LD tensors. As an explicit example, external and
internal components of a HD covariant vector ${\mathbf v}_I$ transform like
\[
{\mathbf v}_\mu\rightarrow J_{\mu}^{\ \kappa}{\mathbf
v}_\kappa+\frac{\partial y^k}{\partial x'^\mu}{\mathbf v}_k\ \ \
\text{and}\ \ \ {\mathbf v}_i\rightarrow  J_{i}^{\ k}{\mathbf v}_k
\]
so that ${\mathbf v}_\mu$ can not be identified with an external
vector, while ${\mathbf v}_i\equiv v_i$ transforms like a LD
internal vector. External and internal components of a HD
contravariant vector ${\mathbf v}^I$ transform according to
\[
{\mathbf v}^\mu\rightarrow {\mathbf v}^\kappa{J^{-1}}_{\kappa}^{\
\mu} \ \ \ \text{and}\ \ \ {\mathbf v}^i\rightarrow {\mathbf
v}^\kappa\frac{\partial y'^i}{\partial x^\kappa}+{\mathbf
v}^k{J^{-1}}_{k}^{\ i}
\]
so that ${\mathbf v}^\mu\equiv v^\mu$ can be identified with a LD
contravariant external vector, while ${\mathbf v}^i$ is not a LD vector.
\\
When constructed from HD tensors, LD tensors are in general functions of
external and internal coordinates. In internal directions the $x^\mu$
dependence just labels the internal space ${\mathrm M}^x_c$ under
consideration. In external directions the $y^i$ dependence will be
eventually removed.

%%%%%%%%%%%%%%%%%%%%%%%%%%%%%%%%%%%%%%%%%%%%%%%%%%%%%%%%%%%%%%%%%
\subsection{\label{sec1.2} Metric}
%%%%%%%%%%%%%%%%%%%%%%%%%%%%%%%%%%%%%%%%%%%%%%%%%%%%%%%%%%%%%%%%%

The most general parameterization of the HD spacetime metric
${\mathbf g}_{IJ}$ covariant under (\ref{STr}) reads
\begin{equation}
{\mathbf g}_{IJ}= \left(
\begin{array}{cc}
g_{\mu\nu}+{h}_{kl}a_\mu^k a_\nu^l & a_\mu^k {h}_{kj}\\
{h}_{il}a_\nu^l & {h}_{ij}
\end{array}
\right) \label{metric}
\end{equation}
with $g_{\mu\nu}(x,y)$, ${h}_{ij}(x,y)$ and $a_\mu^i(x,y)$ functions of
external and internal coordinates that transform according to
\begin{subequations}
\begin{eqnarray}
g_{\mu\nu}&\rightarrow& J_\mu^{\ \kappa}J_\nu^{\
\lambda}g_{\kappa\lambda}\label{g trans}\\
{h}_{ij}&\rightarrow& J_i^{\ k}J_j^{\ l}{h}_{kl}\label{h trans}\\
a_\mu^i&\rightarrow&J_\mu^{\ \kappa}\left(a_\kappa^k {J^{-1}}_k^{\
i} -\partial_\kappa y'^i\right)\label{a trans}
\end{eqnarray}
\end{subequations}
The square matrices $g_{\mu\nu}$ and ${h}_{ij}$ respectively
transform like LD external and internal tensors and can be
identified with metrics on ${\mathrm M}_d$ (after $y^i$ removal)
and ${\mathrm M}^x_c$. The rectangular matrix $a_\mu^i$ transforms
like a LD hybrid tensor up to an inhomogeneous term reminding the
transformation rule of a gauge potential. By means of $a_\mu^i$ it
is also possible to construct a genuine LD hybrid tensor
\begin{equation}
f_{\mu\nu}^i=\partial_\mu a_\nu^i-\partial_\nu a_\mu^i
-a_\mu^j\partial_j a_\nu^i+a_\nu^j\partial_j a_\mu^i \label{f}
\end{equation} appearing as the associated gauge curvature \footnote{
The vanishing of $f_{\mu\nu}^i$ implies the existence of an internal
coordinate transformation setting $a_\mu^i=0$. In general relativity
--identifying space-like coordinates with external variables and time with
the internal coordinate-- the vanishing of $f_{\mu\nu}^i$ characterizes
static gravitational fields.}. It is well known, that this is more than a
similarity in Kaluza-Klein \cite{KK,nonAbKK,KKrev} and embedded spacetime
\cite{gaugeEST} theories, where (\ref{a trans}) precisely
corresponds to the transformation rule of a ${\mathrm
G}^\mathrm{KK}$ or $SO(c)$ gauge potential. On the other hand,
apparently unnoticed is the fact that (\ref{a trans}) always
corresponds to the transformation rule of a vector potential. To
see this explicitly, we read $x$-dependent internal coordinate
transformations (\ref{STr}) as the actions of the internal
diffeomorphism group ${\mathcal Di\!f\!f}_{c}$ on ${\mathrm
M}_c^x$
\begin{equation}
y^i\rightarrow \exp\{\xi^k(x,y)\partial_k\}y^i \nonumber
\end{equation}
with $\xi^k(x,y)$ an appropriate internal vector. By introducing
the operator-valued external covariant vector
\begin{equation}
a_\mu=-i a_\mu^i\partial_i \label{a operator}
\end{equation}
and denoting by ${T} =\exp\{-\xi^k(x,y)\partial_k\}$ the inverse of the
operator generating the transformation, it is straightforward to check  that
(\ref{a trans}) can be rewritten in the familiar gauge transformation form
\begin{equation}
a_\mu\rightarrow {T}a_\mu {T}^{-1}+i{T}(\partial_\mu {T}^{-1})
\label{a gauge trans}
\end{equation}
The off-diagonal term of the HD metric has to be identified with a vector
potential taking values in the internal diffeomorphism algebra of
${di\!f\!f}_{c}$. The associated curvature $f_{\mu\nu}=\partial_\mu a_\nu
-\partial_\nu a_\mu -i[a_\mu,a_\nu]$ corresponds to the operator associated
to $f_{\mu\nu}^i$
\begin{equation}
f_{\mu\nu}=-if_{\mu\nu}^i\partial_i \label{f operator}
\end{equation}
and transforms in the adjoint representation
\begin{equation}
f_{\mu\nu}\rightarrow {T}f_{\mu\nu}{T}^{-1}
\end{equation}
General coordinate transformations do not preserve lengths and
angles, so that the operator $T$ is in general not unitary. The
vanishing of the divergence of $\xi^i$ makes $T$ formally unitary,
a condition always met in Kaluza-Klein and embedded spacetime
theories.
\vskip0,2cm\noindent{\small {\bf Kaluza-Klein}: The HD spacetime
${\mathbf M}_D={\mathrm M}_d\times {\mathcal K}_c$ is the product
manifold of a Lorentz space ${\mathrm M}_d$ and the internal space
${\mathcal K}_c$ admitting an isometry group ${\mathrm
G}^\mathrm{KK}$.  The metric ansatz reads
\begin{equation}
{\mathbf g}_{IJ}= \left(
\begin{array}{cc}
{\rm g}_{\mu\nu}+{\rm A}_\mu^{\sf a}{\rm A}_\nu^{\sf b}{\rm
K}_{\sf a}^k{\rm K}_{\sf b}^l\kappa_{kl}
& {\rm A}_\mu^{\sf a}{\rm K}_{\sf a}^k\kappa_{kj}\\
\kappa_{il}{\rm A}_\nu^{\sf a}{\rm K}_{\sf a}^l & \kappa_{ij}
\end{array}
\right)\tag{\ref{metric}KK} \label{metricKK}
\end{equation}
with ${\rm g}_{\mu\nu}(x)$ a metric on  ${\mathrm M}_d$, $\kappa_{ij}(y)$ a
metric on  ${\mathcal K}_c$, ${\mathrm K}_{\sf a}^k(y)$  Killing vector
fields on ${\mathcal K}_c$ and ${\rm A}_\mu^{\sf a}(x)$ identified with the
gauge potential taking values in the algebra of ${\mathrm G}^\mathrm{KK}$.
By assumption $L_{{\rm K}_{\sf a}}\kappa=0$, equivalently $(\partial_i{\rm
K}_{\sf a}^k)\kappa_{kj}+(\partial_j{\rm K}_{\sf a}^k)\kappa_{ik}+{\rm
K}_{\sf a}^k\partial_k\kappa_{ij}=0$ or $\nabla_i {\rm K}_{{\sf
a}j}+\nabla_j {\rm K}_{{\sf a}i}=0$. Allowed internal coordinate
transformations are generated by Killing vector fields
$\xi^k(x,y)=\epsilon^{\sf a}(x){\rm K}_{\sf a}^k(y)$. Because of the above
identity, $\nabla_i{\rm K}_{\sf a}^i=0$, so that $T$ is unitary. The
transformation rule (\ref{a gauge trans}) yields for ${\rm A}_\mu^{\sf a}$
the ${\mathrm G}^\mathrm{KK}$ gauge potential transformation rule, which
infinitesimally takes the standard form
\begin{equation}
{\rm A}_\mu^{\sf a}\rightarrow {\rm A}_\mu^{\sf a} +{\rm
A}_\mu^{\sf b}\epsilon^{\sf c}c_{\sf bc}^{\sf a} -
\partial_\mu\epsilon^{\sf a}\nonumber
\end{equation}
The corresponding curvature is related to (\ref{f}) by
\begin{equation}
f_{\mu\nu}^i=\left(\partial_\mu {\rm A}_\nu^{\sf c}- \partial_\nu
{\rm A}_\mu^{\sf c}-c_{\sf ab}^{\sf c}{\rm A}_\mu^{\sf a}{\rm
A}_\nu^{\sf b} \right){\rm K}_{\sf c}^i={\rm F}_{\mu\nu}^{\sf
c}{\rm K}_{\sf c}^i \tag{\ref{f}KK} \label{fKK}
\end{equation}}
\vskip0,1cm\noindent{\small {\bf Embedded spacetime}: The HD
spacetime ${\mathbf M}_D\equiv\mathbb{R}^D$ is reduced to a Lorentz space
${\mathrm M}_d$. Denoting be $x^\mu$ the coordinates on ${\mathrm M}_d$, by
${\mathbf t}_\mu$ the associated tangent vectors and by ${\mathbf n}_i(x)$ a
smooth assignment of $c$ orthonormal vectors, ${\mathbf
n}_i\!\cdot\!{\mathbf n}_j=0$, ${\mathbf n}_i\!\cdot\!{\mathbf t}_\mu=0$,
coordinates are adapted by parameterizing internal directions by the
distances $y^i$ along the geodesics leaving ${\mathrm M}_d$ with velocity
${\mathbf n}_i$. In adapted coordinates the flat HD metric reads
\begin{equation}
{\mathbf g}_{IJ}= \left(
\begin{array}{cc}
g_{\mu\nu}+{\rm A}_{\mu m}^{\ \ \ k}{\rm A}_{\mu n}^{\ \ \ l}y^my^n \eta_{kl}&
{\rm A}_{\mu m}^{\ \ \ k}y^m\eta_{kj}\\
\eta_{il} {\rm A}_{\mu n}^{\ \ \ l}y^n & \eta_{ij}
\end{array}
\right)\tag{\ref{metric}{\scriptsize\rm emb}} \label{metricES}
\end{equation}
where $g_{\mu\nu}={\rm g}_{\mu\nu}+2\mathrm{II}_{k\mu\nu}y^k+
\mathrm{II}_{k\mu\kappa} \mathrm{II}_{l\nu}^{\ \ \kappa}y^ky^l$
with ${\rm g}_{\mu\nu}(x)={\mathbf t}_\mu\!\cdot\!{\mathbf t}_\nu$ the
induced metric  and $\mathrm{II}_{i\mu\nu}(x)={\mathbf
t}_\mu\!\cdot\!\partial_\nu{\mathbf n}_i$ the extrinsic curvature (or
\emph{second fundamental form}) of the embedding; $\eta_{ij}$ is a
(pseudo-)Euclidean metric in extra directions; ${\rm A}_{\mu ij}(x)={\mathbf
n}_i\!\cdot\! \partial_\mu{\mathbf n}_j$ is the extrinsic torsion (or
\emph{normal fundamental form}) of the embedding \cite{submanifolds}. The
off-diagonal blocks of (\ref{metricES}) are proportional to the Killing
vectors generating (pseudo-)rotations around the point $y^i=0$ in the flat
internal space. However, the metric is not Kaluza-Klein because of terms
that make $g_{\mu\nu}$ explicitly dependant on $y^i$. Allowed internal
coordinate transformations correspond to the $x$-dependent (pseudo-)rotation
${\mathbf n}_i \rightarrow
\Lambda_i^{\ j}(x){\mathbf n}_j$ and are generated by the Killing
vector fields $\xi^k(x,y)=y^l\omega_l^{\ k}(x)$ with
$\omega_{kl}=-\omega_{lk}$. $\nabla_k y^l\omega_l^{\
k}=\omega_k^{\ k}=0$ so that $T$ is unitary. Under (\ref{STr})
${\rm A}_{\mu i}^{\ \ j}$ transform like a $SO(c)$ gauge potential
\begin{equation}
{\rm A}_{\mu i}^{\ \ j}\rightarrow \Lambda_i^{\ k}{\rm A}_{\mu
k}^{\ \ l}{\Lambda^{-1}}_l^{\ j} -\Lambda_i^{\
k}\partial_\mu{\Lambda^{-1}}_k^{\ j} \nonumber
\end{equation}
The associated curvature is related to (\ref{f}) by
\begin{equation}
f_{\mu\nu}^i=\left(\partial_\mu {\rm A}_{\nu j}^{\ \ \
i}-\partial_\nu {\rm A}_{\mu j}^{\ \ \ i}-[{\rm A}_\mu,{\rm
A}_\nu]_j^{\ i} \right)y^j={\rm F}_{\mu\nu j}^{\ \ \ \ i}y^j
\tag{\ref{f}{\scriptsize\rm emb}} \label{fES}
\end{equation}}
\vskip0,2cm\noindent Denoting by ${\mathbf g}$ the HD metric
determinant and by $g$ and ${h}$ the LD metric determinants, we
have that ${\mathbf g}=gh$. The HD volume element factorizes in
the product of LD volume elements $|{\mathbf g}|^{1/2}=
|g|^{1/2}|{h}|^{1/2}$. The HD inverse metric ${\mathbf g}^{IJ}$
can be evaluated in general terms as
\[
{\mathbf g}^{IJ}= \left(
\begin{array}{cc}
g^{\mu\nu} & -g^{\mu\kappa}a_\kappa^j\\
-a_\lambda^ig^{\lambda\nu} &
{h}^{ij}+a_\kappa^ia_\lambda^jg^{\kappa\lambda}
\end{array}
\right)
\]
with $g^{\mu\nu}$ and ${h}^{ij}$ the inverses of the LD metrics.\\
The parameterization (\ref{metric}) is particularly convenient in connecting
HD with LD geometrical quantities. It generalizes the Kaluza-Klein and
embedded spacetime metric ans\"atze, to the case where no a priori
symmetries or special submanifold have been introduced.

%%%%%%%%%%%%%%%%%%%%%%%%%%%%%%%%%%%%%%%%%%%%%%%%%%%%%%%%%%%%%%%%%%%%%%%%%%%
\subsection{\label{sec1.3}Connections and Curvature Tensors}
%%%%%%%%%%%%%%%%%%%%%%%%%%%%%%%%%%%%%%%%%%%%%%%%%%%%%%%%%%%%%%%%%%%%%%%%%%%

The HD covariant derivative induced by ${\mathbf g}_{IJ}$ is
denoted by $\bm{\nabla}_I$ and acts on tensors as
\[
\bm{\nabla}_I{\mathbf t}_{...J...}^{\ ...}=\partial_I{\mathbf
t}_{...J...}^{\ ...}+ ... -{\mathbf\Gamma}_{IJ}^{\ \ \ K}{\mathbf
t}_{...K...}^{\ ...}+\ ...\
\]
with intrinsic connection coefficients given by ${\mathbf\Gamma}_{IJ}^{\
\ K} = \frac{1}{2}{\mathbf g}^{KL}\left(\partial_I{\mathbf g}_{LJ}+
\partial_J{\mathbf g}_{IL} -\partial_L{\mathbf g}_{IJ}\right)$; by
definition $\bm{\nabla}_K{\mathbf g}_{IJ}=0$ and
$\bm{\nabla}_K|{\mathbf g}|^{1/2}=0$. The commutator of two
covariant derivatives
\[
\left[\bm{\nabla}_I,\bm{\nabla}_J\right]{\mathbf t}_{...K...}^{\
...}=...-{\mathbf R}_{IJK}^{\ \ \ \ L}{\mathbf t}_{...L...}^{\
...}+...
\]
defines the intrinsic curvature tensor
\begin{equation}
{\mathbf R}_{IJK}^{\
\ \ \ L}=\partial_I{\mathbf\Gamma}_{JK}^{\ \ L}-
\partial_J{\mathbf\Gamma}_{IK}^{\ \ L} -{\mathbf\Gamma}_{IK}^{\ \ H}
{\mathbf\Gamma}_{JH}^{\ \ L} + {\mathbf\Gamma}_{JK}^{\ \
H}{\mathbf\Gamma}_{IH}^{\ \ L}\label{HDinR}
\end{equation}
We also denote the Ricci tensor by ${\mathbf R}_{IJ}={\mathbf
R}_{IKJ}^{\ \ \ \ K}$ and the scalar curvature by ${\mathbf
R}={\mathbf g}^{IJ}{\mathbf R}_{IJ}$. The covariant derivative
$\bm{\nabla}_I$ and the associated curvature tensor ${\mathbf
R}_{IJK}^{\ \ \ \ L}$ completely characterize the geometry of the
HD spacetime ${\mathbf M}_D$. We now consider analogous quantities
for the LD internal spaces ${\mathrm M}_c^x$ and external space
${\mathrm M}_d$.
\subsubsection{\label{sec1.3.1}Internal connection and curvatures}
The LD {\em internal covariant derivative} $\nabla_i$ induced by
the metric tensor ${h}_{ij}$
\begin{equation}\nabla_it_{...j...}^{\ ...}=\partial_it_{...j...}^{\
...}+...-{\mathit\Gamma}_{ij}^{\ \ k}t_{...k...}^{\
...}+...\label{IntD}
\end{equation}
with {\em internal intrinsic connection coefficients}
\begin{equation}
{\mathit\Gamma}_{ij}^{\ \ k}
=\frac{1}{2}{h}^{kl}(\partial_i{h}_{lj}+
\partial_j{h}_{il}-\partial_l{h}_{ij})
\end{equation}
is covariant under (\ref{STr}) when acting either on LD internal,
external or hybrid tensors. As a consequence, new LD tensors can
be generated by the action of $\nabla_i$. The commutator
\[
\left[\nabla_i,\nabla_j\right]t_{...k...}^{\ ...}=...-R_{ijk}^{\ \
\ l} t_{...l...}^{\ ...}+...
\]
defines the {\em internal intrinsic curvature}
\begin{equation}
R_{ijk}^{\ \ \ l}= \partial_i{\mathit\Gamma}_{jk}^{\ \ l}-
\partial_j{\mathit\Gamma}_{ik}^{\ \ l}- {\mathit\Gamma}_{ik}^{\ \ m}{\mathit\Gamma}_{jm}^{\ \ l}
+ {\mathit\Gamma}_{jk}^{\ \ m}{\mathit\Gamma}_{im}^{\ \
l}\label{LDinR}
\end{equation}
Internal Ricci tensor and scalar curvature are defined like in
higher dimensions. The internal metric ${h}_{ij}$ and the internal
volume element $|{h}|^{1/2}$ are parallel transported
$\nabla_k{h}_{ij}=0$, $\nabla_k |{h}|^{1/2}=0$. The internal
covariant derivative, however, is not compatible with the external
metric structure as $\nabla_ig_{\mu\nu}\not=0$. External indices
can not be raised, lowered or contracted regardless to the
position of $\nabla_i$. To overcome the problem we extend the
action of $\nabla_i$ to external indices. We define an {\it
internal total covariant derivative} $\nabla^\text{\tiny{tot}}_i$
by
\begin{equation}
\nabla^\text{\tiny{tot}}_it_{...\mu...j...}^{\
...}=\nabla_it_{...\mu... j...}^{\ ...}+...-\hat{E}_{i\mu}^{\ \
\nu} t_{...\nu...j...}^{\ ...} ...\ \label{IntDtot}
\end{equation}
with {\em internal extrinsic connection coefficients} $\hat{E}_{i\mu}^{\ \
\nu}$ chosen so that $\nabla^\text{\tiny{tot}}_kg_{\mu\nu}=0$
(also implying $\nabla^\text{\tiny{tot}}_k |g|^{1/2}=0$). This requirement
fixes the symmetric part of the extrinsic connection to
$\hat{E}_{i(\mu\nu)}=\frac{1}{2}\partial_ig_{\mu\nu}$, living the
antisymmetric part completely arbitrary. It is possible and even natural to
include in $\hat{E}_{i\mu\nu}$ a term proportional to the hybrid tensor
$f_{i\mu\nu}$. Different choices correspond to different internal extrinsic
geometries. In Section \ref{sec2}, equation (\ref{HvLconnectionC}), we will
see that the internal extrinsic connection induced by HD geometry
corresponds to the choice $\hat{E}_{i[\mu\nu]}= \frac{1}{2}f_{i\mu\nu}$. We
therefore set
\begin{equation}
\hat{E}_{i\mu}^{\ \ \nu}=\frac{1}{2}\left(\partial_ig_{\mu\kappa}
+f_{i\mu\kappa}\right)g^{\kappa\nu} \label{IIex}
\end{equation}
Under coordinate redefinitions (\ref{STr}), $\hat{E}_{i\mu}^{\ \
\nu}$ transforms like a genuine LD hybrid tensor
\begin{equation}
\hat{E}_{i\mu}^{\ \ \nu}\rightarrow J_i^{\ j }J_\mu^{\
\kappa}\hat{E}_{j\kappa}^{\ \ \lambda}{J^{-1}}_\lambda^{\
\nu}\nonumber
\end{equation}
\vskip0,2cm\noindent{\small {\bf Kaluza-Klein}: The symmetric part
of the internal extrinsic connection vanishes identically; the
antisymmetric part reduces to the gauge curvature
\begin{equation}
\hat{E}_{i\mu\nu}=\frac{1}{2}{\rm F}_{\mu\nu}^{\sf c}{\rm K}_{{\sf
c}i} \tag{\ref{IIex}KK}\label{IIexKK}
\end{equation}}
\vskip0,05cm \noindent{\small {\bf Embedded spacetime}: The
internal extrinsic connection equals the second fundamental form
$\mathrm{II}_{i\mu\nu}$ of ${\mathrm M}_d$ plus a term linear in
$y^i$
\begin{equation}
\hat{E}_{i\mu\nu}=\mathrm{II}_{i\mu\nu}+\frac{1}{2}\left(
\mathrm{II}_{i\mu\kappa}\mathrm{II}_{j\nu}^{\ \ \kappa}+
\mathrm{II}_{i\nu\kappa}\mathrm{II}_{j\mu}^{\ \ \kappa}- {\rm
F}_{\mu\nu ij}\right)y^j  \tag{\ref{IIex}{\scriptsize\rm
emb}}\label{IIexES}
\end{equation}
On ${\mathrm M}_d$ the linear term vanishes and
$\hat{E}_{i\mu\nu}$ coincides with the second fundamental form
$\hat{E}_{i\mu\nu}|_{y=0} \equiv \mathrm{II}_{i\mu\nu}$.}
\vskip0,2cm\noindent The hybrid tensor $\hat{E}_{i\mu\nu}$ reduces
to the gauge curvature of the external space in Kaluza-Klein backgrounds and
to the extrinsic curvature --second fundamental form-- of the external
spacetime in embedded spacetime models. In Section \ref{sec2} we will see
that $\hat{E}_{i\mu\nu}$ enters the general equations (\ref{HvLRiemann})
relating higher and lower dimensional curvatures in the very same way as the
second fundamental form  enters Gauss, Codazzi and Ricci equations. For
these reasons we will also refer to $\hat{E}_{i\mu\nu}$ as to the {\em
external fundamental form}. The `hat' is introduced to remind us that
${\mathrm M}_d$ is not in general an embedded object and $\hat{E}_{i\mu\nu}$
is not a fundamental form in the standard sense of embedding theory. The
commutator of two total internal covariant derivatives
\begin{eqnarray*}
\left[\nabla^\text{\tiny{tot}}_i,\nabla^\text{\tiny{tot}}_j
\right]t_{...\mu...k...}\!=...-R_{ijk}^{\ \ \
l}t_{...\mu...l...}+...\\  ...-F_{ij\mu}^{\ \ \
\nu}t_{...\nu...k...}+...
\end{eqnarray*}
defines the {\em internal extrinsic curvature}
\begin{equation}
 F_{ij\mu}^{\ \ \ \nu}\!=\partial_i \hat{E}_{j\mu}^{\ \
\nu}-\partial_j\hat{E}_{i\mu}^{\ \ \nu}- \hat{E}_{i\mu}^{\ \
\kappa}\hat{E}_{j\kappa}^{\ \ \nu} + \hat{E}_{j\mu}^{\ \
\kappa}\hat{E}_{i\kappa}^{\ \ \nu}\label{LDinF}
\end{equation}
carrying two internal and two external indices. A direct
computation allows to rewrite $F_{ij\mu\nu}$ as
\begin{equation}
F_{ij\mu\nu}=
\frac{1}{2}\partial_if_{j\mu\nu}-\frac{1}{2}\partial_jf_{i\mu\nu}
+\hat{E}_{i\mu}^{\ \ \kappa}\hat{E}_{j\nu\kappa}-
\hat{E}_{j\mu}^{\ \ \kappa}\hat{E}_{i\nu\kappa}\label{LDinF'}
\end{equation}
\subsubsection{\label{sec1.3.2}External connection and curvatures}
The definition of covariant differentiation along external direction is less
straightforward. The derivative $\nabla_\mu$ associated with the external
metric $g_{\mu\nu}$ is not a covariant LD object. Difficulties already
emerge at the scalar level. The allowed external coordinate dependence of
internal coordinate redefinitions produces an inhomogeneous term in the
transformation rule of partial derivatives
\begin{equation}
\partial_\mu\rightarrow\partial'_\mu=J_\mu^{\
\nu}\left(\partial_\nu+\frac{\partial y^i}{\partial
x^\nu}\partial_i\right) \nonumber
\end{equation}
The problem can be resolved by adding a counter term proportional to
$a_\mu^i$ which also transform inhomogeneously.  The derivative operator
\begin{equation}
\hat\partial_\mu=\partial_\mu-i a_\mu \label{hat-d}
\end{equation}
transforms like a genuine LD external vector when acting on
scalars
\begin{equation}
\hat\partial_\mu\rightarrow\hat\partial'_\mu= J_\mu^{\
\nu}\hat\partial_\nu \nonumber
\end{equation}
On the other hand, the commutator of two hatted derivatives is no
longer vanishing
\[
\left[\hat\partial_\mu,\hat\partial_\nu\right]=-if_{\mu\nu}
\]
Differentiation is extended to LD external tensors by introducing
the generalized Christoffel symbols
\begin{equation}
{\mathit{\hat\Gamma}}_{\mu\nu}^{\ \
\kappa}=\frac{1}{2}g^{\kappa\lambda}(\hat\partial_\mu
g_{\lambda\nu}+\hat\partial_\nu
g_{\mu\lambda}-\hat\partial_\lambda g_{\mu\nu})
\end{equation}
where ordinary derivatives are replaced by hatted ones in the
standard definition. Generalized Christoffel symbols transform
like proper connection symbols. The {\em external covariant
derivative} $\hat\nabla_\mu$
\begin{equation}
\hat\nabla_\mu t_{...\nu...}^{\ ...}=\hat\partial_\mu
t_{...\nu...}^{\ ...}+... - {\mathit{\hat\Gamma}}_{\mu\nu}^{\ \
\kappa}t_{...\kappa...}^{\ ...}
\end{equation}
is covariant under ($\ref{STr}$) when acting on LD external
tensors. New LD tensors can be generated by the action of
$\hat\nabla_\mu$ on external tensors. The commutator
\[
\left[\hat\nabla_\mu,\hat\nabla_\nu\right]t_{...\kappa...}^{\
...}=...-\hat{R}_{\mu\nu\kappa}^{\ \ \
\lambda}t_{...\lambda...}^{\ ...}+...
-f_{\mu\nu}^i\nabla^\text{\tiny{tot}}_it_{...\kappa...}^{\ ...}
\]
defines a genuine {\em  external intrinsic curvature} tensor as
\begin{eqnarray}
\hat{R}_{\mu\nu\kappa}^{\ \ \ \lambda}=
\hat\partial_\mu{\mathit{\hat\Gamma}}_{\nu\kappa}^{\ \ \lambda}
-\hat\partial_\nu{\mathit{\hat\Gamma}}_{\mu\kappa}^{\ \ \lambda}+\hskip2,5cm\nonumber\\
-{\mathit{\hat\Gamma}}_{\mu\kappa}^{\ \
\rho}{\mathit{\hat\Gamma}}_{\nu\rho}^{\ \
\lambda}+{\mathit{\hat\Gamma}}_{\nu\kappa}^{\ \
\rho}{\mathit{\hat\Gamma}}_{\mu\rho}^{\ \ \lambda}+
f_{\mu\nu}^i\hat{E}_{i\kappa}^{\ \ \lambda} \label{LDexR}
\end{eqnarray}
External Ricci and scalar curvatures are defined as usual by
contraction $\hat{R}_{\mu\nu}=\hat{R}_{\mu\kappa\nu}^{\ \ \
\kappa}$ and $\hat{R}=g^{\mu\nu}\hat{R}_{\mu\nu}$. It is worth
noticing that $\hat{R}_{\mu\nu\kappa}^{\ \ \ \lambda}$,
$\hat{R}_{\mu\nu}$ and $\hat{R}$ are reducible tensors.
\vskip0,2cm\noindent{\small {\bf Kaluza-Klein}: In Kaluza-Klein
theories $\hat{R}$ does not correspond with the scalar curvature
${\mathrm R}$ associate with the four dimensional metric ${\rm
g}_{\mu\nu}(x)$. Equation (\ref{LDexR}) yields
\begin{equation}
\hat{R}={\mathrm R}+{\rm F}^{\sf a}_{\mu\nu} {\rm F}^{{\sf
a}\mu\nu}/2 \tag{\ref{LDexR}KK}\label{LDexRKK}
\end{equation}
with gauge indices contracted with the group metric.}
\vskip0,05cm \noindent{\small {\bf Embedded spacetime}: The
corresponding equation in embedded spacetime theories is  more complicated
involving, apart from the gauge field ${\rm F}_{\mu\nu i}^{\ \ \ \ j}$, the
external fundamental forms $\hat{E}_{i\mu\nu}$. Specializing to $y^i=0$ we
obtain
\begin{equation}
\hat{R}={\mathrm R}+ {\mathcal
O}(y)\tag{\ref{LDexR}{\scriptsize\rm emb}}\label{LDexRES}
\end{equation}
with ${\mathrm R}$ the intrinsic curvature associated with the
metric induced on the submanifold.}
\vskip0,2cm\noindent The external metric $g_{\mu\nu}$ and the
external volume element $|g|^{1/2}$ are parallel transported
$\hat\nabla_\kappa g_{\mu\nu}=0$, $\hat\nabla_\kappa|g|^{1/2}=0$. On the
other hand, it is not even possible to ask wether the external derivative is
compatible with internal metric structures, because $\hat\nabla_\mu$ is not
covariant when acting on internal and hybrid tensors. Both problems can be
resolved by extending the action of $\hat\nabla_\mu$ to internal indices. We
define the {\it external total covariant derivative}
$\hat\nabla^\text{\tiny{tot}}_\mu$ by
\begin{equation}
\hat\nabla^\text{\tiny{tot}}_\mu t_{...\kappa...
k...}=\hat\nabla_\mu t_{...\kappa... k...}+...-\hat{C}_{\mu k}^{\
\ l}t_{...\kappa... l...}+...\label{ExtDtot}
\end{equation}
where the {\em external extrinsic connection coefficients}
$\hat{C}_{\mu k}^{\ \ l}$ are determined by the requirement of
covariance and by the compatibility condition
$\hat\nabla^\text{\tiny{tot}}_\mu{h}_{ij}=0$ (also implying
$\hat\nabla^\text{\tiny{tot}}_\mu|{h}|^{1/2}=0$). We obtain
\begin{equation}
\hat{C}_{\mu k}^{\ \ l}=\partial_k a_\mu^l +{E}_{\mu km}{h}^{ml}
\label{Ain}
\end{equation}
 where
\begin{equation}
{E}_{\mu ij}=\frac{1}{2}[\hat\partial_\mu{h}_{ij}- (\partial_i
a_\mu^k){h}_{kj}-(\partial_j a_\mu^k){h}_{ik}] \label{IIin}
\end{equation}
transforms like a genuine LD hybrid tensor.  At any given external
point ${E}_{\mu ij}|_x$ corresponds to the standard second
fundamental form  describing the embedding of ${\mathrm M}^x_c$ in
${\mathbf M}_D$. ${E}_{\mu ij}$ generalizes  the notion of second
fundamental form to the whole foliation of the HD spacetime in
internal spaces. For this reason we refer to ${E}_{\mu ij}$ as the
{\em internal fundamental form}. For later use we also rewrite
${E}_{\mu ij}$ as
\begin{equation}
{E}_{\mu ij}=\frac{1}{2}(\partial_\mu h_{ij}-\nabla_ia_{\mu
j}-\nabla_ja_{\mu i})\label{IIinKK'}
\end{equation}
and note that  the following identity holds
\begin{equation}
\hat\nabla^\text{\tiny{tot}}_\mu {E}_{\nu ij}-
\hat\nabla^\text{\tiny{tot}}_\nu {E}_{\mu ij}=
\frac{1}{2}\left(\nabla_if_{j\nu\mu}+\nabla_jf_{i\nu\mu}\right)
\label{IIinIdentity}
\end{equation}
Under the residual general covariance group (\ref{STr}) external
extrinsic connection coefficients transform like a genuine $GL(c)$
connection
\begin{equation}
\hat{C}_{\mu k}^{\ \ l}\rightarrow J_\mu^{\ \nu}(J_k^{\ m}
\hat{C}_{\nu m}^{\ \ n} {J^{-1}}_n^{\ l}-J_{k}^{\
m}\hat\partial_\nu {J^{-1}}_{m}^{\ l}) \nonumber
\end{equation}
\vskip0,2cm\noindent{\small {\bf Kaluza-Klein}: By virtue of the
identity $(\partial_i{\rm K}_{\sf a}^k)\kappa_{kj}+(\partial_j{\rm
K}_{\sf a}^k)\kappa_{ik}+{\rm K}_{\sf a}^k\partial_k\kappa_{ij}=0$
the internal fundamental form  vanishes identically
\begin{equation}
{E}_{\mu ij}=0 \tag{\ref{IIin}KK}\label{IIinKK}
\end{equation}
The embedding of each ${\mathcal K}_c$ is {\em totally geodesic}
\cite{submanifolds}. The external extrinsic connection
coefficients only depends on off-diagonal blocks of the metric
\begin{equation}
\hat{C}_{\mu k}^{\ \ l}= {\rm A}_\mu^{\sf a} (\partial_k {\rm
K}_{\sf a}^l) \tag{\ref{Ain}KK}
\end{equation} }
\vskip0,05cm \noindent{\small {\bf Embedded spacetime}: Since the
internal metric $\eta_{ij}$ does not depend on coordinates and
$A_{\mu kl}$ are antisymmetric in internal indices the embedding
of internal spaces is again {\em totally geodesic}
\begin{equation}
{E}_{\mu ij}=0 \tag{\ref{IIin}{\scriptsize\rm emb}}\label{IIinES}
\end{equation}
The external extrinsic connection reduces to the normal
fundamental form of the embedding
\begin{equation}
\hat{C}_{\mu k}^{\ \ l}=  {\rm A}_{\mu k}^{\ \
l}\tag{\ref{Ain}{\scriptsize\rm emb}}
\end{equation}}
\vskip0,2cm\noindent The commutator of two external total
covariant derivatives yields the associated curvature forms
\begin{eqnarray*}
\left[\hat\nabla^\text{\tiny{tot}}_\mu,\hat\nabla^\text{\tiny{tot}}_\nu
\right]t_{...\kappa... k...}=...-\hat{R}_{\mu\nu\kappa}^{\ \ \
\lambda}t_{...\lambda... k...}+\hskip1cm\\ +... - \hat{F}_{\mu\nu
k }^{\ \ \ l}t_{...\kappa... l...}+
...-f_{\mu\nu}^i\nabla^\text{\tiny{tot}}_i t_{...\kappa... k...}
\end{eqnarray*}
where the {\em external extrinsic curvature} tensor, carrying two
external and two internal indices, is defined as
\begin{eqnarray}
\hat{F}_{\mu\nu k }^{\ \ \ l}=\hat\partial_\mu\hat{C}_{\nu k}^{\ \
l}-\hat\partial_\nu \hat{C}_{\mu k}^{\ \ l}
+\hskip2,5cm\nonumber\\
-\hat{C}_{\mu k}^{\ \ m}\hat{C}_{\nu m}^{\ \ l}+\hat{C}_{\nu k}^{\
\ m}\hat{C}_{\mu m}^{\ \ l}+f_{\mu\nu}^i{\mathit\Gamma}_{ik}^{\ \
l}\label{LDexF}
\end{eqnarray}
With the help of (\ref{IIinIdentity}) a straightforward
computation allows to evaluate $\hat{F}_{\mu\nu k }^{\ \ \ l}$
directly in terms of $f_{\mu\nu}^i$ and ${E}_{\mu ij}$ as
\begin{equation}
\hat{F}_{\mu\nu kl}=
\frac{1}{2}\partial_kf_{l\mu\nu}-\frac{1}{2}\partial_lf_{k\mu\nu}+
{E}_{\mu k}^{\ \ i}{E}_{\nu li}- {E}_{\nu k}^{\ \ i}{E}_{\mu
li}\label{LDexF'}
\end{equation}
a formula that closely resembles (\ref{LDinF'}).

\subsubsection{\label{sec1.3.3} Hybrid curvatures}
The commutator of external and internal total covariant derivatives defines
one more curvature tensor that describes the tangling of ${\mathrm M}_d$ and
${\mathrm M}_c^x$ in ${\mathbf M}_D$
\begin{eqnarray*}
\left[\hat\nabla^\text{\tiny{tot}}_\mu,\nabla^\text{\tiny{tot}}_i\right]
t_{...\kappa... k...}=...-\hat{H}_{\mu i\kappa}^{\ \ \
\lambda}t_{...\lambda... k...}+\hskip0.4cm  \\ +... -H_{\mu i k
}^{\ \ \ l}t_{...\kappa... l...}+\hskip0.2cm
\\ +...+\hat{E}_{i\mu}^{\ \ \nu} \hat\nabla^\text{\tiny{tot}}_\nu
t_{...\kappa... k...}- {E}_{\mu i}^{\ \
j}\nabla^\text{\tiny{tot}}_j t_{...\kappa... k...}
\end{eqnarray*}
where the two {\em hybrid curvature} tensors $\hat{H}_{\mu
i\kappa}^{\ \ \ \lambda}$ and $H_{\mu ik}^{\ \ \ l}$ have the form
\begin{eqnarray}
\hat{H}_{\mu i\kappa}^{\ \ \
\lambda}=\hat\partial_\mu\hat{E}_{i\kappa}^{\ \
\lambda}-\partial_i{\mathit{\hat\Gamma}}_{\mu\kappa}^{\
\ \lambda}+\hskip2.5cm\nonumber\\
-{\mathit{\hat\Gamma}}_{\mu\kappa}^{\ \ \nu}\hat{E}_{i\nu}^{\ \
\lambda} +\hat{E}_{i\kappa}^{\ \
\nu}{\mathit{\hat\Gamma}}_{\mu\nu}^{\ \ \lambda}
-(\partial_ia_\mu^j)\hat{E}_{j\kappa}^{\ \ \lambda}\label{LDhyH}
\end{eqnarray}
and
\begin{eqnarray}
H_{\mu ik}^{\ \ \ l}=\hat\partial_\mu{\mathit\Gamma}_{ik}^{\
\ l} -\partial_i\hat{C}_{\mu k}^{\ \ l}+\hskip2.5cm \nonumber\\
- \hat{C}_{\mu k}^{\ \ j}{\mathit\Gamma}_{ij}^{\ \ l}
+{\mathit\Gamma}_{ik}^{\ \ j}\hat{C}_{\mu j}^{\ \ l}
-(\partial_ia_\mu^j){\mathit\Gamma}_{jk}^{\ \ l}\label{LDhyHh}
\end{eqnarray}
A direct computation allows to reexpress the hybrid curvatures in
terms of the sole  fundamental forms $\hat{E}_{i\mu\nu}$ and
${{E}}_{\mu ij}$ as
\begin{eqnarray}
\hat{H}_{\mu i\kappa\lambda}=
\hat\nabla^\text{\tiny{tot}}_\lambda\hat{E}_{i\kappa\mu}-
\hat\nabla^\text{\tiny{tot}}_\kappa\hat{E}_{i\lambda\mu}+\hskip2.0cm\nonumber\\
+{E}_{\lambda i}^{\ \ k}\hat{E}_{k\mu\kappa}+ {E}_{\kappa k}^{\ \
k}\hat{E}_{k\mu\lambda}+ f^j_{\kappa\lambda}{E}_{\mu
ik}\label{LDhyH'}
\end{eqnarray}
\begin{eqnarray}
H_{\mu ikl}= \nabla^\text{\tiny{tot}}_k{E}_{\mu li}
-\nabla^\text{\tiny{tot}}_l{E}_{\mu ki}+\hskip2.0cm\nonumber\\
+\hat{E}_{k\mu}^{\ \ \nu}{E}_{\nu li}- \hat{E}_{l\mu}^{\ \
\nu}{E}_{\nu ki}\label{LDhyHh'}
\end{eqnarray}
Therefore, the four LD tensors $\hat{R}_{\mu\nu\kappa}^{\ \ \ \lambda}$,
$R_{ijk}^{\ \ \ l}$, ${{E}}_{\mu ij}$, $\hat{E}_{i\mu\nu}$ give a complete
characterization of the intrinsic and extrinsic geometry of external and
internal spaces. Note that $f_{i\mu\nu}$ is the antisymmetric part of
$\hat{E}_{i\mu\nu}$ and $a_\mu^i$ is related to it by (\ref{f}). It is
curious that in spite of the different role played by external and internal
coordinates the formalism is symmetric under their interchange.
 The symmetry is substantial only when $f^i_{\mu\nu}\equiv0$ and
${\mathbf M}_D$ double foliates in internal and external
directions.

%%%%%%%%%%%%%%%%%%%%%%%%%%%%%%%%%%%%%%%%%%%%%%%%%%%%%%%%%%%%%%%%
\subsection{\label{sec1.4} Reference Frames}
%%%%%%%%%%%%%%%%%%%%%%%%%%%%%%%%%%%%%%%%%%%%%%%%%%%%%%%%%%%%%%%%
Besides standard tensor calculus in holonomic coordinates, there
is a second formalism that allows to successfully deal with
geometrical problems: the tetrad (in four dimensions) or reference
frame formalism. Among other things, it allows to clarify the role
of gauge invariance for the gravitational field \cite{UtiKib} and
is indispensable to deal with general relativistic interactions of
spinors. In this section we show that the reference frame
formalism is also the natural language to deal with
dimensional reduction problems.\\
In the HD spacetime, we consider pseudo-orthogonal covariant and
contravariant {\em reference frames} ${\mathbf r}_I^{\ A}$ and
${\mathbf r}_A^{\ I}$, decomposing the metric and its inverse as
${\mathbf g}_{IJ}={\mathbf r}_I^{\ A}{\mathbf r}_J^{\
B}\eta_{AB}$, ${\mathbf g}^{IJ}={\mathbf r}_A^{\ I}{\mathbf
r}_B^{\ J}\eta^{AB}$. In terms of the metric parametrization
(\ref{metric})
\begin{equation}
{\mathbf r}_I^{\ A}= \left(
\begin{array}{cc}
r_\mu^{\ \alpha} & a_\mu^k\rho_k^{\ a}\\
0 & \rho_i^{\ a}
\end{array}
\right)\ \ {\mathbf r}_A^{\ I}= \left(
\begin{array}{cc}
r_\alpha^{\ \mu} & -r_\alpha^{\ \kappa}a_\kappa^i\\
0 & \rho_a^{\ i}
\end{array}
\right)\label{rf}
\end{equation}
with $r_\mu^{\ \alpha}$, $r_\alpha^{\ \mu}$  and $\rho_i^{\ a}$,
$\rho_a^{\ i}$ decomposing the LD metrics, $r_\mu^{\
\alpha}r_\nu^{\ \beta}\eta_{\alpha\beta}=g_{\mu\nu}$, $\rho_i^{\
a}\rho_j^{\ b}\eta_{ab}={h}_{ij}$ etc. Reference vectors are determined up
to point dependent pseudo-rotations expressing observer's freedom of
arbitrarily choosing the reference frame. Hence, reference frames transform
as holonomic vectors under general coordinate transformations and like
pseudo-Euclidean vectors under pseudo-rotations. The theory is covariant
under
\[
{\mathbf r}_A^{\ I}\rightarrow {\mathbf r}_A^{\ J}{J^{-1}}_J^{\
I}\text{,}\ \ \ {\mathbf r}_A^{\ I}\rightarrow
{\mathbf\Lambda}_A^{\ B}{\mathbf r}_B^{\ I}
\]
with ${\mathbf\Lambda}_A^{\ B}({\mathbf x})$ any point dependent,
pseudo-orthogonal matrix, ${\mathbf\Lambda}_A^{\ C}
{\mathbf\Lambda}_B^{\ D}\eta_{CD}=\eta_{AB}$. When coordinate
invariance is broken, local pseudo-orthogonal transformations get
restricted to the block diagonal form
\begin{equation}
{\mathbf\Lambda}_A^{\ B}(\bm{\mathrm x})= \left(
\begin{array}{cc}
\Lambda_\alpha^{\ \beta}(x)& 0 \\
0& \Lambda_a^{\ b}(x,y)
\end{array}\right)\nonumber
\end{equation}
with $\Lambda_\alpha^{\ \beta}(x)$ and $\Lambda_a^{\ b}(x,y)$
lower dimensional pseudo-orthogonal matrices: $\Lambda_\alpha^{\
\gamma}\Lambda_\alpha^{\
\delta}\eta_{\gamma\delta}=\eta_{\alpha\beta}$ and $\Lambda_a^{\
c}\Lambda_b^{\ d}\eta_{cd}=\eta_{ab}$. The LD vectors $r_\alpha^{\
\mu}$ and $\rho_a^{\ i}$ correctly transform as LD reference
frames
\begin{subequations}
\begin{eqnarray*}
r_\alpha^{\ \mu}&\rightarrow& r_\alpha^{\
\kappa}{J^{-1}}_\kappa^{\ \mu} \text{,}\ \ r_\alpha^{\
\mu}\rightarrow \Lambda_\alpha^{\ \beta} r_\beta^{\ \mu}\\
\rho_a^{\ i}&\rightarrow& \rho_a^{\ k}{J^{-1}}_k^{\ i}
 \text{,}\ \ \  \rho_a^{\ i}\rightarrow\Lambda_a^{\
b}\rho_b^{\ i}
\end{eqnarray*}
\end{subequations}
We fix the following notation for Kaluza-Klein and embedded
spacetime models
\vskip0,2cm\noindent{\small {\bf Kaluza-Klein}: LD reference
frames are denoted by
\begin{equation}
r_\mu^{\ \alpha} ={\rm r}_\mu^{\ \alpha}(x)\hskip0,5cm \rho_i^{\
a}=k_i^{\ a}(y) \tag{\ref{rf}KK}\label{rfKK}
\end{equation}
with ${\rm g}_{\mu\nu}={\rm r}_\mu^{\ \alpha} {\rm r}_\nu^{\
\beta}\eta_{\alpha\beta}$ and $\kappa_{ij}= k_i^{\ a}k_j^{\
b}\eta_{ab}$.}
\vskip0,05cm \noindent{\small {\bf Embedded spacetime}: LD
reference frames are chosen as
\begin{equation}
r_\mu^{\ \alpha} =(\delta_\mu^\kappa+y^i\mathrm{II}_{i\mu}^{\ \
\kappa}){\rm t}_\kappa^{\ \alpha}(x) \hskip0,5cm \rho_i^{\ a}={\rm
n}_i^{\ a}(x) \tag{\ref{rf}{\scriptsize\rm emb}}\label{rfES}
\end{equation}
with ${\rm g}_{\mu\nu}={\rm t}_\mu^{\ \alpha}{\rm t}_\nu^{\
\beta}\eta_{\alpha\beta}$ and $\eta_{ij}= {\rm n}_i^{\ a}{\rm
n}_j^{\ b}\eta_{ab}$.}
%
%%%%%%%%%%%%%%%%%%%%%%%%%%%%%%%%%%%%%%%%%%%%%%%%%%%%%%%%%%%%%%
\subsection{\label{sec1.5} More on Tensors}
%%%%%%%%%%%%%%%%%%%%%%%%%%%%%%%%%%%%%%%%%%%%%%%%%%%%%%%%%%%%%%
 Instead of specifying HD tensors by giving their components with
respect to the holonomic coordinate system, we can specify them by
giving their projections on the reference frame
\[
{\mathbf t}_{...A...}^{...B...}=...{\mathbf r}_{A}^{\ I}...\
{\mathbf t}_{...I...}^{...J...}\ ...{\mathbf r}_{J}^{\ B}...
\]
These quantities are invariant under general coordinate transformations and
transform like pseudo-Euclidean tensor components under point dependent
reference frame redefinition
\[
{\mathbf t}_{...A...}^{...B...}\rightarrow
...{\mathbf\Lambda}_{A}^{\ C} ...\ {\mathbf
t}_{...C...}^{...D...}\ ...{{\mathbf\Lambda}^{-1}}_{D}^{\ B}...
\]
LD external, internal and hybrid tensor components
$t_{...\alpha...}^{...\beta...}$, $t_{...a...}^{...b...}$ and $t_{...\alpha
... a... }^{...\beta...b...}$ are introduced with analogous conventions and
transformation properties
\begin{eqnarray*}
t_{...\alpha...}^{...\beta...} \!&\rightarrow&\!
...\Lambda_{\alpha}^{\
\gamma}...\ t_{...\gamma...}^{...\delta...}\ ...{\Lambda^{-1}}_{\delta}^{\ \beta}...\\
t_{...a...}^{...b...} \!&\rightarrow& \!...\Lambda_{a}^{\ c}...\
t_{...c...}^{...d...}\ ...{\Lambda^{-1}}_{d}^{\ b}...\\
t_{...\alpha...a...}^{...\beta...b...}\!&\rightarrow&\!...\Lambda_{\alpha}^{\
\gamma}...\Lambda_{a}^{\ c}...\
t_{...\gamma...c...}^{...\delta...d...}...\
{\Lambda^{-1}}_{\delta}^{\ \beta}...{\Lambda^{-1}}_{d}^{\ b}...
\end{eqnarray*}
 It is readily checked that, when HD covariance is broken {\em
pseudo-orthogonal components of HD tensors transform like
(pseudo-)orthogonal components of LD tensors.} For example, external and
internal components of a HD covariant vector ${\mathbf v}_A={\mathbf r}_A^{\
I}{\mathbf v}_I$ transform like
\[
{\mathbf v}_\alpha\rightarrow \Lambda_\alpha^{\ \beta}{\mathbf
v}_\beta\ \ \ \text{and}\ \ \  {\mathbf v}_a\rightarrow
\Lambda_{a}^{\ b}{\mathbf v}_b
\]
so that $v_\alpha\equiv{\mathbf v}_\alpha$ and $v_a\equiv{\mathbf
v}_a$ may be identified with the components of two LD external and
internal vectors. A HD rank-two covariant tensor ${\mathbf
b}_{AB}$ produces an external $b_{\alpha\beta}\equiv{\mathbf
b}_{\alpha\beta}$, an internal $b_{ab}\equiv{\mathbf b}_{ab}$ and
two hybrid $b_{\alpha b}\equiv{\mathbf b}_{\alpha b}$, $b'_{\alpha
b}\equiv{\mathbf b}_{b\alpha}$ LD rank-two covariant tensors. This
makes the use of pseudo-orthogonal reference frames particularly
convenient in investigating dimensional reduction problems.

%%%%%%%%%%%%%%%%%%%%%%%%%%%%%%%%%%%%%%%%%%%%%%%%%%%%%%%%%%%%%
\subsection{\label{sec1.6} More on Connections and Curvature Tensors}
%%%%%%%%%%%%%%%%%%%%%%%%%%%%%%%%%%%%%%%%%%%%%%%%%%%%%%%%%%%%%

The whole machinery of calculus on manifolds is readily transposed in the
reference frame formalism by defining a covariant derivative acting on both,
curved and flat spacetime indices
\begin{equation} {\mathbf D}_I{\mathbf
t}_{...A...}=\bm{\nabla}_I{\mathbf t}_{...A...}+...-{\mathbf
\Omega}_{I,A}^{\ \ \ B}{\mathbf t}_{...B...}+...
\end{equation}
with connection coefficients
${\mathbf\Omega}_{I,AB}=(\bm{\nabla}_I {\mathbf r}_A^{\ K})
{\mathbf r}_B^{\ L}{\mathbf g}_{KL}$. With these conventions
${\mathbf D}_I{\mathbf r}_A^{\ J}\equiv0$. The commutator of two
covariant derivatives yields the intrinsic curvature tensor
\begin{eqnarray}
{\mathbf R}_{IJAB}=\partial_I{\mathbf\Omega}_{J,AB}-
\partial_J{\mathbf\Omega}_{I,AB}+\hskip1,5cm\nonumber\\
-{\mathbf\Omega}_{I,A}^{\ \ \ C}{\mathbf\Omega}_{J,CB}
+{\mathbf\Omega}_{J,A}^{\ \ \ C}{\mathbf\Omega}_{I,CB}
\label{HDRrf}
\end{eqnarray}
which is related to (\ref{HDinR}) by  contraction with reference frames,
${\mathbf R}_{IJKL}={\mathbf R}_{IJAB}{\mathbf r}_K^{\ A} {\mathbf r}_L^{\
B}$. In LD internal and external spaces we proceed along the very same
lines.
\subsubsection{\label{sec1.6.1} Internal connection and curvatures}
On internal spaces, we define an {\em internal total covariant
derivative} $D^\text{\tiny{tot}}_i$ as
\begin{eqnarray}
D^\text{\tiny{tot}}_it_{...\alpha... a...}=
\nabla^\text{\tiny{tot}}_it_{...\alpha... a...}+ ...-
{\mathit\Omega}_{i,a}^{\ \ \ b}t_{...\alpha... b...}+\nonumber
\\+...- {A}_{i,\alpha}^{\ \ \ \beta}t_{...\beta... a...}+...
\end{eqnarray}
with connection coefficients ${\mathit\Omega}_{i,ab}=
(\nabla^\text{\tiny{tot}}_i\rho_a^{\ k})\rho_b^{\ l}{h}_{kl}$ and
${A}_{i,\alpha\beta}= (\nabla^\text{\tiny{tot}}_ir_\alpha^{\
\kappa})r_\beta^{\ \lambda}g_{\kappa\lambda}$. Under coordinate
redefinitions ${\mathit\Omega}_{i,ab}$ and ${A}_{i,\alpha\beta}$ transform
like genuine internal tensors. Under local (pseudo-)rotations of reference
frames, ${\mathit\Omega}_{i,ab}$ transforms like an $SO(c)$ gauge connection
while ${A}_{i,\alpha\beta}$ behave like a  tensor
\begin{eqnarray}
{\mathit{\Omega}}_{i,a}^{\ \ \ b}&\rightarrow& \Lambda_a^{\
c}{\mathit{\Omega}}_{i,c}^{\ \ \ d} \Lambda_d^{\ b}-\Lambda_a^{\
c}(\partial_i \Lambda_c^{\ b})\\
{A}_{i,\alpha}^{\ \ \ \beta} & \rightarrow&\Lambda_\alpha^{\
\gamma}{A}_{i,\gamma}^{\ \ \ \delta} \Lambda_\delta^{\
\beta}\label{A trans}
\end{eqnarray}
With these conventions $D^\text{\tiny{tot}}_i\rho_a^{\ j}=0$ and
$D^\text{\tiny{tot}}_ir_\alpha^{\ \mu}=0$. The commutator of two
total internal covariant derivatives yields the intrinsic and
extrinsic curvature tensors
\begin{equation}
R_{ijab}=
\partial_i{\mathit\Omega}_{j,ab}-\partial_j{\mathit\Omega}_{i,ab}- {\mathit\Omega}_{i,a}^{\ \
\ c}{\mathit\Omega}_{j,cb}+{\mathit\Omega}_{j,a}^{\ \ \
c}{\mathit\Omega}_{i,cb} \label{LDinRrf}
\end{equation}
and
\begin{equation}
{F}_{ij\alpha\beta}=
\partial_i{A}_{j,\alpha\beta}-\partial_j{A}_{i,\alpha\beta}-
{A}_{i,\alpha}^{\ \ \ \gamma}{A}_{j,\gamma\beta}+
{A}_{j,\alpha}^{\ \ \ \gamma}{A}_{i,\gamma\beta} \label{LDinFrf}
\end{equation}
which are related to (\ref{LDinR}) and (\ref{LDinF}) by
contraction with LD reference frames,
$R_{ijkl}={R}_{ijab}\rho_k^{\ a}\rho_l^{\ b}$ and
$F_{ij\kappa\lambda}= {F}_{ij\alpha\beta}r_\kappa^{\
\alpha}r_\lambda^{\ \beta}$.
\subsubsection{\label{sec1.6.2}External connection and curvatures}
On the external space, we define an {\em external total covariant
derivative} $\hat{D}^\text{\tiny{tot}}_\mu$ as
\begin{eqnarray}
\hat{D}^\text{\tiny{tot}}_\mu t_{...\alpha... a...}=
\hat\nabla^\text{\tiny{tot}}_\mu t_{...\alpha... a...}+
   ...- {\mathit{\hat\Omega}}_{\mu,\alpha}^{\ \ \ \beta} t_{...\beta... a...}+
\nonumber\\+...- \hat{A}_{\mu,a}^{\ \ \ b}t_{...\alpha...b...}+...
\end{eqnarray}
with connection coefficients ${\mathit{\hat\Omega}}_{\mu,
\alpha\beta}= (\hat\nabla^\text{\tiny{tot}}_\mu r_\alpha^{\
\kappa})r_\beta^{\ \lambda}g_{\kappa\lambda}$ and
$\hat{A}_{\mu,ab} =(\hat\nabla^\text{\tiny{tot}}_\mu\rho_a^{\ k})\rho_b^{\
l}{h}_{kl}$. Under coordinate transformations
${\mathit{\hat\Omega}}_{\mu,\alpha\beta}$ and $\hat{A}_{\mu,ab}$ behaves
like genuine external tensors. Under local redefinition of reference frames
${\mathit{\hat\Omega}}_{\mu,\alpha\beta}$ and $\hat{A}_{\mu,ab}$ transform
as $SO(d)$ and $SO(c)$ connections respectively
\begin{eqnarray}
{\mathit{\hat\Omega}}_{\mu,\alpha}^{\ \ \ \beta}&\rightarrow&
\Lambda_\alpha^{\ \gamma}{\mathit{\hat\Omega}}_{\mu,\gamma}^{\ \ \
\delta} \Lambda_\delta^{\ \beta}-\Lambda_\alpha^{\
\gamma}(\hat\partial_\mu
 \Lambda_\gamma^{\ \beta})\\
\hat{A}_{\mu,a}^{\ \ \ b}&\rightarrow&\Lambda_a^{\
c}\hat{A}_{\mu,c}^{\ \ \ d} \Lambda_d^{\ b}-\Lambda_a^{\
c}(\hat\partial_\mu
 \Lambda_c^{\ b})\label{hatA gauge trans}
\end{eqnarray}
As above $\hat\nabla^\text{\tiny{tot}}_\mu r_\alpha^{\ \nu}=0$ and
$\hat\nabla^\text{\tiny{tot}}_\mu\rho_a^{\ i}=0$. The commutator
of two external total covariant derivative again yields the
intrinsic and extrinsic curvature tensors
\begin{eqnarray}
\hat{R}_{\mu\nu\alpha\beta}=
\hat\partial_\mu{\mathit{\hat\Omega}}_{\nu,\alpha\beta}
-\hat\partial_\nu{\mathit{\hat\Omega}}_{\mu,\alpha\beta}+\hskip2,5cm\nonumber
\\-{\mathit{\hat\Omega}}_{\mu,\alpha}^{\ \ \
\gamma}{\mathit{\hat\Omega}}_{\nu,\gamma\beta}+{\mathit{\hat\Omega}}_{\nu,\alpha}^{\
\ \ \gamma}{\mathit{\hat\Omega}}_{\mu,\gamma\beta}
+f_{\mu\nu}^i{A}_{i,\alpha\beta} \label{LDexRrf}
\end{eqnarray}
and
\begin{eqnarray}
\hat{F}_{\mu\nu ab}=\hat\partial_\mu\hat{A}_{\nu,ab}-
\hat\partial_\nu\hat{A}_{\mu,ab}+\hskip2,5cm\nonumber\\
- \hat{A}_{\mu,a}^{\ \ \ c}\hat{A}_{\nu,cb}+ \hat{A}_{\nu,a}^{\ \
\ c}\hat{A}_{\mu,cb}+f_{\mu\nu}^i{\mathit\Omega}_{i,ab}
\label{LDexFrf}
\end{eqnarray}
again related to (\ref{LDexR}) and (\ref{LDexF}) by contraction
with LD reference frames,
$\hat{R}_{\mu\nu\kappa\lambda}=\hat{R}_{\mu\nu\alpha\beta}r_\kappa^{\
\alpha}r_\lambda^{\ \beta}$ and $\hat{F}_{\mu\nu
kl}=\hat{F}_{\mu\nu ab} \rho_k^{\ a}\rho_l^{\ b}$.

\subsubsection{\label{sec2.6.3}Hybrid curvatures}
The commutator of total external and internal derivative yields
the hybrid curvatures
\begin{eqnarray}
\hat{H}_{\mu i\alpha\beta}= \hat\partial_\mu A_{i,\alpha\beta}-
\partial_i{\mathit{\hat\Omega}}_{\mu,\alpha\beta}+\hskip2,5cm\nonumber\\
-{\mathit{\hat\Omega}}_{\mu,\alpha}^{\ \ \
\gamma}A_{i,\gamma\beta}+ A_{i,\alpha}^{\ \ \
\!\gamma}{\mathit{\hat\Omega}}_{\mu,\gamma\beta}-
(\partial_ia_\mu^j)A_{j,\alpha\beta} \label{LDHrf}
\end{eqnarray}
and
\begin{eqnarray}
H_{\mu iab}= \hat\partial_\mu{\mathit\Omega}_{i,ab}-
\partial_i\hat{A}_{\mu,ab}+\hskip2,5cm\nonumber\\
-\hat{A}_{\mu,a}^{\ \ \ c}{\mathit\Omega}_{i,cb}+
{\mathit\Omega}_{i,a}^{\ \ \ \!c}\hat{A}_{\mu,cb}-
(\partial_ia_\mu^j){\mathit\Omega}_{j,ab} \label{LDHhrf}
\end{eqnarray}
related to (\ref{LDhyH}) and (\ref{LDhyHh}) by contraction with LD
reference frames and that can be rewritten in terms of the
pseudo-orthogonal components of  fundamental forms
\begin{eqnarray}
{E}_{\gamma ab}=r_\gamma^{\ \kappa}\rho_a^{\ i} \rho_b^{\
j}{E}_{\kappa ij}, \hskip0,5cm \hat{E}_{c \alpha\beta}=\rho_c^{\
k}r_\alpha^{\ \mu}r_\beta^{\ \nu}\hat{E}_{k\mu\nu} \label{IIrf}
\end{eqnarray}
Nothing has really changed; the pseudo-Euclidean tensors
$\hat{R}_{\alpha\beta\gamma\delta}$, $R_{abcd}$, ${E}_{\gamma ab}$ and
$\hat{E}_{c \alpha\beta}$ completely characterize the geometry of
dimensional reduction.

%%%%%%%%%%%%%%%%%%%%%%%%%%%%%%%%%%%%%%%%%%%%%%%%%%%%%%%%%%%%%%%
%
\section{\label{sec2} Reducing Geometry}
%
%%%%%%%%%%%%%%%%%%%%%%%%%%%%%%%%%%%%%%%%%%%%%%%%%%%%%%%%%%%%%%%

We are now in position to write down general equations that relate the
higher and lower dimensional geometries. In holonomic coordinates this task
requires very long and tedious calculations with results that are not always
transparent. Instead, within the reference frames formalism, it is almost
straightforward to establish the desired relations. The formulas obtained in
this section extend and unify well known identities of Kaluza-Klein and
submanifold theories.

%%%%%%%%%%%%%%%%%%%%%%%%%%%%%%%%%%%%%%%%%%%%%%%%%%%%%%%%%%%%%%%
\subsection{\label{sec2.1} Connection coefficients}
%%%%%%%%%%%%%%%%%%%%%%%%%%%%%%%%%%%%%%%%%%%%%%%%%%%%%%%%%%%%%%%
\begin{subequations}\label{HvLconnection}
In the reference frames formalism, HD connection coefficients
directly relate to LD intrinsic connection coefficients,
fundamental forms and extrinsic connection coefficients in the
following way
\begin{equation}
{\mathbf r}_\gamma^{\ I}{\mathbf\Omega}_{I,\alpha\beta}=
r_\gamma^{\ \mu}{\mathit{\hat\Omega}}_{\mu,\alpha\beta}\\
\label{HvLconnectionA}
\end{equation}
\begin{equation}
{\mathbf r}_c^{\ I}{\mathbf\Omega}_{I,ab}= \rho_c^{\ i}
{\mathit\Omega}_{i,ab} \label{HvLconnectionB}
\end{equation}
\begin{equation}
{\mathbf r}_\gamma^{\ I}{\mathbf\Omega}_{I,a\beta}=
\hat{E}_{a\gamma\beta}\label{HvLconnectionC}
\end{equation}
\begin{equation}
{\mathbf r}_c^{\ I}{\mathbf\Omega}_{I,\alpha b}= {E}_{\alpha cb}
\label{HvLconnectionD}
\end{equation}
\begin{equation}
{\mathbf r}_\gamma^{\ I}{\mathbf\Omega}_{I,ab}= r_\gamma^{\
\mu}\hat{A}_{\mu,ab} \label{HvLconnectionE}
\end{equation}
\begin{equation}
{\mathbf r}_c^{\ I}{\mathbf\Omega}_{I,\alpha\beta}= \rho_c^{\ i}
A_{i,\alpha\beta}\label{HvLconnectionF}
\end{equation}
Analogous equations connecting HD Christoffel symbols with LD quantities are
much more complicated. By means of relations (\ref{HvLconnection}) it is
straightforward to relate HD to LD curvatures, geodesic equations and
geometric operators.

\end{subequations}

%%%%%%%%%%%%%%%%%%%%%%%%%%%%%%%%%%%%%%%%%%%%%%%%%%%%%%%%%%%%%%%
\subsection{\label{sec2.2} Riemann Curvatures: \protect\\
 extension of  Gauss, Codazzi and Ricci equations}
%%%%%%%%%%%%%%%%%%%%%%%%%%%%%%%%%%%%%%%%%%%%%%%%%%%%%%%%%%%%%%%

Gauss, Codazzi and Ricci equations give relations between HD
curvature and LD curvatures, second and normal fundamental forms
of a submanifold and provide, at the same time, integrability
conditions for a subspace to be embeddable in a HD
spacetime~\cite{submanifolds}. They are important in a variety of
physical applications, especially in general relativity. Recently,
they have been extended to foliations and applied to the analysis
of embedded spacetimes \cite{GCR}. Equations of an apparently
different nature relating HD  curvature to LD curvatures and gauge
fields are also the key ingredient of Kaluza-Klein unification
schemes \cite{KK,nonAbKK,KKrev}. Both set of equations are special
cases of the general equations relating the HD Riemann tensor
${\mathbf R}_{ABCD}$ to LD Riemann tensors $\hat{R}_{\alpha\beta
\gamma\delta}$, $R_{abcd}$  and  fundamental forms ${E}_{\gamma
ab}$, $\hat{E}_{c\alpha\beta}$. The symmetries of the Riemann
tensor allow only six independent projections on external/internal
directions.
\begin{subequations}\label{HvLRiemann}
\subsubsection{Gauss type equations}

The external components of the HD Riemann tensor are related to the external
intrinsic curvature and fundamental forms by an equation which is formally
identical to the Gauss equation for an embedded space
\begin{equation}
{\mathbf R}_{\alpha\beta\gamma\delta}=
\hat{R}_{\alpha\beta\gamma\delta}+ \hat{E}_{a\alpha\gamma}
\hat{E}^a_{\ \beta\delta}- \hat{E}_{a\beta\gamma} \hat{E}^a_{\
\alpha\delta} \label{GaussEX}
\end{equation}
In spite of this analogy it is worth remarking that the external space
${\mathrm M}_d$ is not an embedded object, $\hat{R}_{\alpha\beta
\gamma\delta}$ is not a standard Riemannian curvature tensor and
$\hat{E}_{c \alpha\beta}$ has an antisymmetric part keeping truck of the
gauge field $f^i_{\mu\nu}$. The internal components of the HD Riemann tensor
are related to the internal intrinsic curvature and the fundamental forms
yielding again an equation formally identical to the Gauss equation for an
embedded space
\begin{equation}
{\mathbf R}_{abcd}= R_{abcd}+{E}_{\alpha ac}{E}^{\alpha}_{\ bd}
-{E}_{\alpha bc}{E}^{\alpha }_{\ ad}\label{GaussIN}
\end{equation}
This time the analogy is more than formal. For every given value $x^\mu$ of
the external coordinates  the internal space ${\mathrm M}_c^x$ is an
embedded object in ${\mathbf M}_D$. In this case, $R_{abcd}|_x$ is the
relative Riemann tensor and ${E}_{\gamma ab}|_x$ is the second fundamental
form so that (\ref{GaussIN}) correspond to a genuine Gauss equation for the
embedding.
\subsubsection{Codazzi type equations}

HD Riemann tensor components with three indices of one sort and one index of
the other, are related to the LD hybrid curvatures and the fundamental
forms. These terms yield the generalization of the Codazzi equation for the
external space ${\mathrm M}_d$
\begin{equation}
{\mathbf R}_{\alpha b\gamma\delta}= \hat{H}_{\alpha b\gamma\delta}
+{E}_{\gamma b}^{\ \ a}\hat{E}_{a\alpha\delta} -{E}_{\delta b}^{\
\ a}\hat{E}_{a\alpha\gamma} \label{CodazziEX}
\end{equation}
and for the foliation of ${\mathbf M}_D$ in the internal spaces
${\mathrm M}_c^x$
\begin{equation}
{\mathbf R}_{\alpha bcd}=H_{\alpha bcd} -{\hat{E}}_{c \alpha}^{\ \
\alpha}{E}_{\alpha bd} +{\hat{E}}_{d \alpha}^{\ \
\alpha}{E}_{\alpha bc} \label{CodazziIN}
\end{equation}
The explicit appearance of the hybrid curvatures $\hat{H}_{\alpha
b\gamma\delta}$ and $H_{\alpha bcd}$ can be eliminated by means of
(\ref{LDhyH'}) and (\ref{LDhyHh'}), giving the Codazzi equations in their
more familiar form
\begin{equation}
{\mathbf R}_{\alpha b\gamma\delta}=
\hat{D}^\text{\tiny{tot}}_\delta\hat{E}_{b \gamma\alpha}-
\hat{D}^\text{\tiny{tot}}_\gamma\hat{E}_{b \delta\alpha}+
f^i_{\gamma\delta}{E}_{\alpha ib}
\tag{\ref{CodazziEX}$'$}\label{CodazziEX'}
\end{equation}
\begin{equation}
{\mathbf R}_{\alpha bcd}= D^\text{\tiny{tot}}_c{E}_{\alpha db}-
D^\text{\tiny{tot}}_d{E}_{\alpha cb}
\tag{\ref{CodazziIN}$'$}\label{CodazziIN'}
\end{equation}
The interpretation of these equations requires the same caution
used for generalized Gauss equations. While (\ref{CodazziIN'}) are
genuine Codazzi equations for the embedded spaces ${\mathrm
M}_c^x$, (\ref{CodazziEX'}) correspond to  standard Codazzi
equations only when $f^i_{\mu\nu}=0$ and the external space
${\mathrm M}_d$ reduce to an embedded object.
\subsubsection{Ricci type equations}

HD Riemann tensor components with the first two indices of one sort and the
last two indices of the other, relate the LD extrinsic curvatures
(\ref{LDexF}), (\ref{LDinF}) to the hybrid tensor (\ref{f}), yielding a
single equation
\begin{eqnarray}
{\mathbf R}_{\alpha\beta cd}=\hat{F}_{\alpha\beta
cd}+F_{cd\alpha\beta}+\hskip2.0cm\nonumber\\
-\frac{1}{2}r_\alpha^{\ \mu}r_\beta^{\ \nu}\rho_c^{\ k}\rho_d^{\
l} \left(\partial_kf_{l\mu\nu}-\partial_lf_{k\mu\nu}\right)
\label{Ricci}
\end{eqnarray}
This generalizes the Ricci equation for both, the external space ${\mathrm
M}_d$ and the foliation in internal subspaces ${\mathrm M}_c^x$. The
explicit appearance of the external extrinsic curvature
$\hat{F}_{\alpha\beta cd}$ or of the internal extrinsic curvature
$F_{cd\alpha\beta}$ (or of both of them), can be removed by means of
(\ref{LDexF'}) and (\ref{LDinF'}). It respectively yields the standard form
of the Ricci equation for the external space ${\mathrm M}_d$
\begin{equation}
{\mathbf R}_{\alpha\beta cd}=\hat{F}_{\alpha\beta cd}+
\hat{E}_{c\alpha}^{\ \ \gamma}\hat{E}_{d\beta\gamma}
-\hat{E}_{c\beta}^{\ \ \gamma}\hat{E}_{d\alpha\gamma}
\tag{\ref{Ricci}$'$}\label{RicciEX}
\end{equation}
and for the foliation in internal spaces ${\mathrm M}_c^x$
\begin{equation}
{\mathbf R}_{\alpha\beta cd}=F_{cd\alpha\beta}+ {E}_{\alpha c}^{\
\ a}{E}_{\beta da}- {E}_{\beta c}^{\ \ a}{E}_{\alpha da}
\tag{\ref{Ricci}$''$}\label{RicciIN}
\end{equation}
Once again, a little caution in the interpretation of
(\ref{Ricci}), or (\ref{RicciEX}), or (\ref{RicciIN}), is
necessary.
\subsubsection{The sixth equation}

The remaining group of HD Riemann tensor components relates the fundamental
forms to their total covariant derivatives, yielding an equation that has no
equivalent in the theory of embedding
\begin{eqnarray}
{\mathbf R}_{\alpha b \gamma d}=
\hat{D}^\text{\tiny{tot}}_\alpha{E}_{\gamma bd}
+D^\text{\tiny{tot}}_b\hat{E}_{d\alpha\gamma}+\hskip2.0cm\nonumber\\
+{E}_{\alpha b}^{\ \ a}{E}_{\gamma ad}+ \hat{E}_{b\alpha}^{\ \
\beta}\hat{E}_{d\beta\gamma} \label{sixth}
\end{eqnarray}
This equation appears as a further integrability condition for the
tangling of ${\mathrm M}_d$ and ${\mathrm M}_c^x$ in ${\mathbf
M}_D$ and consists a new result obtained by this approach.
\end{subequations}

%%%%%%%%%%%%%%%%%%%%%%%%%%%%%%%%%%%%%%%%%%%%%%%%%%%%%%%%%%%%%%%
\subsection{\label{sec2.3} Ricci curvatures}
%%%%%%%%%%%%%%%%%%%%%%%%%%%%%%%%%%%%%%%%%%%%%%%%%%%%%%%%%%%%%%%
\begin{subequations}
By contracting the generalized Gauss, Codazzi, Ricci equations and
(\ref{sixth}) we easily obtain the external
\begin{eqnarray}
{\mathbf R}_{\alpha\beta}=\hat{R}_{\alpha\beta}+
\hat{D}^\text{\tiny{tot}}_\alpha{E}_{\beta c}^{\ \ c}+{E}_{\alpha
cd}{E}_\beta^{\
dc}+\nonumber\\
+D^\text{\tiny{tot}}_c\hat{E}_{\ \alpha\beta}^{c}
+\hat{E}_{c\alpha\beta}\hat{E}_{\ \gamma}^{c\ \gamma}
\end{eqnarray}
hybrid
\begin{eqnarray}
{\mathbf R}_{\alpha b}=\hat{D}^\text{\tiny{tot}}_\alpha
\hat{E}_{b\gamma}^{\ \ \gamma}- \hat{D}^\text{\tiny{tot}}_\gamma
\hat{E}_{b\alpha}^{\ \
\gamma}-f_{\alpha\gamma}^c\hat{E}_{\ cb}^\gamma+\nonumber\\
+D^\text{\tiny{tot}}_b{E}_{\alpha c}^{\ \ c}-
D^\text{\tiny{tot}}_c{E}_{\alpha b}^{\ \ c}
\end{eqnarray}
and internal
\begin{eqnarray}
{\mathbf R}_{ab}=R_{ab}+\hat{D}^\text{\tiny{tot}}_\gamma {E}_{\
ab}^\gamma+{E}_{\gamma ab}
{E}_{\ c}^{\gamma\ c}+\nonumber\\
+D^\text{\tiny{tot}}_a\hat{E}_{b\gamma}^{\ \ \gamma}
+\hat{E}_{a\gamma\delta}\hat{E}_{b}^{\ \delta\gamma}
\end{eqnarray}
components of the HD Ricci tensor. From the viewpoint of pure higher
dimensional gravity these equations display the most general kind of LD
matter that can be obtained in induced-matter theories
\cite{STM}. \label{HvLRicci}
\end{subequations}

%%%%%%%%%%%%%%%%%%%%%%%%%%%%%%%%%%%%%%%%%%%%%%%%%%%%%%%%%%%%%%%
\subsection{\label{sec2.4} Scalar curvatures}
%%%%%%%%%%%%%%%%%%%%%%%%%%%%%%%%%%%%%%%%%%%%%%%%%%%%%%%%%%%%%%%
The eventual contraction of equations (\ref{HvLRicci}) yields the
identity connecting the HD scalar curvature with LD intrinsic and
extrinsic curvatures, lying at the heart of Lagrangian reduction
of HD Einstein gravity. We display it in standard tensor formalism
\begin{eqnarray}
{\mathbf R}= \hat{R}+2\nabla_i\hat{E}_{\ \mu}^{i\ \mu}+\hat{E}_{i
\mu}^{\ \ \mu}\hat{E}_{\ \nu}^{i\
\nu}+\hat{E}_{i\mu\nu}\hat{E}^{i\nu\mu}+\nonumber\\
+R+2\hat{\nabla}_\mu{E}_{\ i}^{\mu\ i}+{E}_{\mu i}^{\ \ i} {E}_{\
j}^{\mu\ j}+{E}_{\mu ij}{E}^{\mu ji} \label{HvLscalar}
\end{eqnarray}
This equation generalizes well known relations holding in
Kaluza-Klein and submanifold theories.
\vskip0,2cm\noindent{\small {\bf Kaluza-Klein}: In virtue of
(\ref{IIexKK}), (\ref{LDexRKK}) and (\ref{IIinKK}) equation
(\ref{HvLscalar}) reduces to
\begin{equation}
{\mathbf R}={\mathrm R}+R+\frac{1}{4} {\rm F}^{\sf a}_{\mu\nu}
{\rm F}^{{\sf a}\mu\nu} \tag{\ref{HvLscalar}KK}
\end{equation}
with ${\mathrm R}$ the standard scalar curvature associated with
the four dimensional metric ${\rm g}_{\mu\nu}(x)$.}
\vskip0,05cm \noindent{\small {\bf Embedded spacetime}: By
recalling (\ref{IIexES}), (\ref{LDexRES}), (\ref{IIinES}) and the
fact that we are considering spacetimes embedded in flat HD
spacetime  equation (\ref{HvLscalar}) evaluated at $y^i=0$
reproduces the well known identity
\begin{equation}
{\mathrm R}+\mathrm{II}_{i \mu}^{\ \ \mu}\mathrm{II}_{\ \nu}^{i\
\nu}-\mathrm{II}_{i\mu\nu}\mathrm{II}^{i\mu\nu}=0
\tag{\ref{HvLscalar}{\scriptsize\rm emb}}
\end{equation}
that relates intrinsic and extrinsic curvature scalars for a submanifold
embedded in a HD flat space.}
\vskip0,2cm\noindent By means of equations
(\ref{GaussEX})-(\ref{sixth}), (\ref{HvLRicci}) and (\ref{HvLscalar})  it is
possible to obtain general reduction formulas for the Weyl conformal tensor,
which also plays an important role in the analysis of dimensional
reduction~\cite{Jackiw06}.

%%%%%%%%%%%%%%%%%%%%%%%%%%%%%%%%%%%%%%%%%%%%%%%%%%%%%%%%%%%%%%%
\subsection{\label{sec2.5} Geodesic motion}
%%%%%%%%%%%%%%%%%%%%%%%%%%%%%%%%%%%%%%%%%%%%%%%%%%%%%%%%%%%%%%%%
Free motion in  HD spacetime  is described by geodesic equations
\begin{equation}
\ddot{\mathbf x}^K+{\mathbf\Gamma}^K_{IJ}\dot{\mathbf
x}^I\dot{\mathbf x}^J =0\nonumber
\end{equation}
where $\dot{\mathbf x}^I=d\mathbf x^I/d\tau$ is the HD velocity vector. As
discussed in Section \ref{sec1.1} external contravariant components of HD
vectors behave like  LD vectors, so that $\dot{\mathbf x}^\mu=\dot{x}^\mu$
is identified with the LD external velocity. On the other hand internal
contravariant components do not, so that $\dot{\mathbf x}^i=\dot{y}^i$ is
not a LD object. The definition of a LD vector once again involves $a_\mu^i$
\[
\hat{\dot{y}}^i=\dot{y}^i+a^i_\mu \dot{x}^\mu
\]
HD geodesic equations split in two groups that separately transform under
the residual covariance group (\ref{STr}). The first group describes a no
longer free motion in external directions and its coupling to internal
variables through the fundamental forms, is given by
\begin{subequations}
\begin{equation}
\ddot{x}^\kappa+{\hat\Gamma}^\kappa_{\mu\nu}\dot{x}^\mu\dot{x}^\nu+2\hat{E}_{i\
 \mu}^{\ \kappa}\hat{\dot{y}}^i \dot{x}^\mu -
 {E}_{\ ij}^{\kappa}\hat{\dot{y}}^i \hat{\dot{y}}^j
=0 \label{geoex}
\end{equation}
The second group takes in to account internal motion and its
dynamical interaction with external variables
\begin{equation}
\dot{\hat{\dot{y}}}^k+\Gamma^k_{ij}\hat{\dot{y}}^i\hat{\dot{y}}^j
+(\partial_ia_\mu^k)\dot{x}^\mu \hat{\dot{y}}^i+ 2{E}_{\mu\ i}^{\
k}\dot{x}^\mu \hat{\dot{y}}^i -\hat{E}_{\ \mu\nu}^{k}\dot{x}^\mu
\dot{x}^\nu=0 \label{geoin}
\end{equation}
(the first three terms of the left hand side can be recast in the LD
covariant expression $\dot{x}^\mu\hat\nabla^\text{\tiny{tot}}_\mu
\hat{\dot{y}}^k+\hat{\dot{y}}^i\nabla^\text{\tiny{tot}}_i
\hat{\dot{y}}^k-{E}_{\mu\ i}^{\ \ \!k}
\hat{\dot{y}}^i\dot{x}^\mu$). The interaction between internal and
external motion vanishes if and only if the fundamental forms
identically vanish, $\hat{E}_{i\mu\nu}=0$ and $\mathit{E}_{\mu
ij}=0$. Specializing to Kaluza-Klein and embedded spacetime models
we obtain: \label{geo}
\end{subequations}
\vskip0,2cm\noindent{\small {\bf Kaluza-Klein}: Taking into
account (\ref{IIexKK}), (\ref{IIinKK}) equations (\ref{geo})
reduce to
\begin{equation}
\begin{array}{l}
\ddot{x}^\kappa+\Gamma^\kappa_{\mu\nu}\dot{x}^\mu\dot{x}^\nu+ {\rm
q}_{\sf a}{\rm F}_{\
\!\mu}^{{\sf a}\ \!\kappa}\dot{x}^\mu=0\\
\dot{{\rm q}}_{\sf a}-c_{\sf ab}^{\sf c}\dot{x}^\mu{\rm A}^{\sf
b}_\mu {\rm q}_{\sf c}=0
\end{array}
\tag{\ref{geo}KK}
\end{equation}
where ${\rm q}_{\sf a}={\rm K}_{{\sf a}i}(y)\hat{\dot{y}}^i$. The first
equation describes the external motion of a particle of vector charge ${\rm
q}_{\sf a}$ in the possibly non-Abelian gauge field ${\rm F}_{\
\!\mu\nu}^{\sf a}$. The second equation describes the rotation of the
charge-vector in the group space. In the case of a one dimensional Abelian
group ${\rm q}_{\sf 1}$ is constant in time and the first equation reduces
to the classical Lorentz equation of a charged particle moving on a manifold
in an electromagnetic field.}
\vskip0,05cm \noindent{\small {\bf Embedded spacetime}: In a
neighborhood of radius $\epsilon$ of a submanifold the equations
(\ref{IIexES}), (\ref{IIinES}) allow to rewrite (\ref{geo}) in the form
\begin{equation}
\begin{array}{l}
\ddot{x}^\kappa+\Gamma^\kappa_{\mu\nu}\dot{x}^\mu\dot{x}^\nu+
\frac{1}{2}{\rm L}_{ij}{\rm F}_{\mu}^{\ \kappa ij}\dot{x}^\mu
+2\hat{\dot{y}}^i\mathrm{II}_{i\mu}^{\ \ \!\kappa}\dot{x}^\mu+
{\cal O}(\epsilon)=0\\
\dot{{\rm L}}^{ij}-\dot{x}^\mu{\rm A}_{\mu\ k}^{\ \ \! i}{\rm
L}^{kj} +\dot{x}^\mu{\rm A}_{\mu\ k}^{\ \ \! j}{\rm L}^{ki} +{\cal
O}(\epsilon)=0
\end{array}\tag{\ref{geo}{\scriptsize\rm emb}}
\end{equation}
where ${\rm L}^{ij}=y^i\hat{\dot{y}}^j-y^j\hat{\dot{y}}^i$ is the angular
momentum in internal directions and $\mathrm{II}_{i\mu\nu}(x)$ the second
fundamental form of the embedding. Higher order terms in $\epsilon$ can be
neglected only if some physical mechanism constrains the system  in a
sufficiently small neighborhood of the submanifold. As in standard
Kaluza-Klein theories the first equation describes the external motion of a
particle of charge $\frac{1}{2}{\rm L}_{ij}$ in the gauge field ${\rm
F}_{\mu\nu}^{\ \ \ ij}$; the non-trivial dependence of external metric on
internal coordinates produces the extra term
$2\hat{\dot{y}}^i\mathrm{II}_{i\mu}^{\ \
\!\kappa}\dot{x}^\mu$ making geodesics to drift away from the
submanifold. The second equation describes the precession of
internal angular momentum produced by the extrinsic torsion of the
embedding.}
 \vskip0,2cm \noindent Equations (\ref{geo}) can also
be obtained from the Lagrangian
\begin{equation}
{\mathcal L}=\frac{1}{2}g_{\mu\nu}\dot{x}^\mu\dot{x}^\nu
+\frac{1}{2}h_{ij}\left(\dot{y}^i+a^i_\mu \dot{x}^\mu\right)
\left(\dot{y}^j+a^j_\nu \dot{x}^\nu\right)
\end{equation}
For later considerations it is also useful to write the
corresponding Hamiltonian
\begin{equation}
{\mathcal H}=\frac{1}{2}g^{\mu\nu}\left(p_\mu-a_\mu^i\pi_i\right)
\left(p_\nu-a_\nu^j\pi_j\right)+\frac{1}{2}h^{ij}\pi_i\pi_j
\label{H}
\end{equation}
with $p_\mu=\partial{\mathcal L}/\partial\dot{x}^\mu$,
$\pi_i=\partial{\mathcal L}/\partial\dot{y}^i$ the momenta conjugated to
external and internal coordinates respectively. Internal momenta $\pi_i$
correctly transform as LD vectors, while LD external covariant momenta have
to be defined as $\hat{p}_\mu\equiv p_\mu-a_\mu^i\pi_i$.

%%%%%%%%%%%%%%%%%%%%%%%%%%%%%%%%%%%%%%%%%%%%%%%%%%%%%%%%%%%%%%%
\subsection{\label{sec2.6}Geometric operators}
%%%%%%%%%%%%%%%%%%%%%%%%%%%%%%%%%%%%%%%%%%%%%%%%%%%%%%%%%%%%%%%%
 We now consider the dimensional reduction of Laplace
and Dirac operators.
\subsubsection{Laplace operator}
In every local coordinate frame the HD scalar Laplace operator
${\mathbf\Delta}$ takes the form
\begin{equation}
{\mathbf\Delta} =|{\mathbf g}|^{-1/2}\partial_I {\mathbf
g}^{IJ}|{\mathbf g}|^{1/2}\partial_J\nonumber
\end{equation}
${\mathbf\Delta}$ is Hermitian with respect to the standard scalar
product constructed by means of the HD covariant measure
$|{\mathbf g}|^{1/2}d{\mathbf x}=|g|^{1/2}|h|^{1/2}dxdy$. By
rewriting the operator in terms of covariant derivatives,
recalling the inverse metric decomposition and the relations
(\ref{HvLconnection}) between HD and LD connection coefficients,
we  obtain the most general decomposition covariant under
(\ref{STr})
\begin{eqnarray}
{\mathbf\Delta} =\nonumber|g|^{-1/2} \left(\hat\partial_\mu\!+\!
\frac{1}{2}{E}_{\mu i}^{\ \ i} \right)
g^{\mu\nu}|g|^{1/2}\left(\hat\partial_\nu\!+\! \frac{1}{2}{E}_{\nu
i}^{\ \ i}\right)\!+ \nonumber\\
\!+|{h}|^{-1/2}\left(\partial_i\! +\! \frac{1}{2}\hat{E}_{i\mu}^{\
\ \mu}\right) {h}^{ij}|{h}|^{1/2}\left(\partial_j\!+\!
\frac{1}{2}\hat{E}_{j\mu}^{\ \
\mu}\right)\!+ \nonumber\\
\!-\!\frac{1}{2}\hat\nabla_\mu{E}_{\ i}^{\mu\ i}
\!-\!\frac{1}{4}{E}_{\mu i}^{\ \ i}{E}_{\ j}^{\mu\
j}\!-\!\frac{1}{2}\nabla_i\hat{E}_{\ \mu}^{i\ \mu}\!-\!
\frac{1}{4}\hat{E}_{i\mu}^{\ \ \mu} \hat{E}_{\ \nu}^{i\
\nu}\hskip0,2cm \label{LapRecuction}
\end{eqnarray}
The first righthand side term of this equation corresponds to the
LD external Laplace operator
\begin{equation}
\Delta^{\rm ext} =|g|^{-1/2}\partial_\mu
g^{\mu\nu}|g|^{1/2}\partial_\nu\nonumber
\end{equation}
(Hermitian with respect to the external scalar product constructed
by means of the LD volume element $|g|^{1/2}dx$) with partial
derivatives $\partial_\mu$ replaced by the HD Hermitian operators
\[
\hat\partial_\mu+\frac{1}{2}{E}_{\mu i}^{\ \
i}=\partial_\mu+\left(\partial_\mu\ln
|h|^{1/4}\right)-ia_\mu-\frac{1}{2}\nabla_ia_\mu^i
\]
The total derivative $\partial_\mu\ln |h|^{1/4}$ takes into account the
different normalization of HD and LD states. It amounts to the rescaling
$\Delta^{\rm ext}\rightarrow|{h}|^{-1/4}
\Delta^{\rm ext}|{h}|^{1/4}$. The Hermitian internal operator
$a_\mu-\frac{i}{2}\nabla_ia_\mu^i$ enters the expression as a
gauge potential.  The second righthand term of
(\ref{LapRecuction}) corresponds to the LD internal Laplace
operator
\begin{equation}
\Delta^{\rm int}=
|{h}|^{-1/2}\partial_i{h}^{ij}|{h}|^{1/2}\partial_j \nonumber
\end{equation}
(Hermitian with respect to the internal scalar product constructed
by means of the LD volume element $|h|^{1/2}dy$) with partial
derivatives $\partial_i$ replaced by
\[
\partial_i+
\frac{1}{2}\hat{E}_{i\mu}^{\ \ \mu}=\partial_i+\left(\partial_i\ln
|g|^{1/4}\right)
\]
As above, the total derivative amounts to the rescaling
$\Delta^{\rm int}\rightarrow |g|^{-1/4}\Delta^{\rm int}
|g|^{1/4}$, necessary to correct the different normalization of HD
and LD states. The remaining terms in the righthand side of
(\ref{LapRecuction}) are identified with a scalar potential
induced by dimensional reduction. They are know to produce
observable effects in low energy physics \cite{gaugeEST}.
Specializing to Kaluza-Klein and embedded spacetime models we
obtain:
\vskip0,2cm\noindent{\small {\bf Kaluza-Klein}: By recalling
(\ref{metricKK}), (\ref{IIexKK}), (\ref{LDexRKK}), (\ref{IIinKK}),
the fact that $\nabla_i{\rm K}_{\sf a}^i=0$ and assuming that the
external metric only depends on external coordinates,
(\ref{LapRecuction}) reduces to the well known expression
\begin{equation}
\begin{array}{l}
{\mathbf\Delta}^{\rm KK}= |{\rm g}|^{-1/2}\left(\partial_\mu-
i{\rm A}_\mu^{\sf a}\hat{\rm K}_{\sf a}\right){\rm g}^{\mu\nu}
|{\rm g}|^{1/2}\left(\partial_\nu-i {\rm
A}_\nu^{\sf a}\hat{\rm K}_{\sf a}\right)+\\
\hskip1.2cm+|\kappa|^{-1/2}\partial_i\kappa^{ij}|\kappa|^{1/2}\partial_j
\end{array}\tag{\ref{LapRecuction}KK}\label{Laplace KK}
\end{equation}
where $\hat{\rm K}_{\sf a}=-i{\rm K}_{\sf a}^i\partial_i$ are
infinite dimensional Hermitian generators of the isometry
algebra.}
\vskip0,05cm \noindent{\small {\bf Embedded spacetime}: By
recalling (\ref{metricES}), (\ref{IIexES}), (\ref{LDexRES}),
(\ref{IIinES}), the fact that $\nabla_i{\rm A}_j^{\ i}y^j={\rm
A}_i^{\ i}=0$ and after rescaling fields and operators by
\begin{equation}
\begin{array}{rcl}
\bm\Psi&\rightarrow& |g|^{1/4}|{\rm g}|^{-1/4}\bm\Psi \\
{\mathbf\Delta} &\rightarrow& |g|^{1/4} |{\rm
g}|^{-1/4}{\mathbf\Delta}\ |g|^{-1/4}|{\rm g}|^{1/4}
\end{array}\nonumber
\end{equation}
in a neighborhood of radius $\epsilon$ of ${\mathrm M}_d$
(\ref{LapRecuction}) reduces to
\begin{equation}
\begin{array}{l}
{\mathbf\Delta}^{\rm emb}= |{\rm g}|^{-1/2}
\left(\partial_\mu-\frac{i}{2}{\rm A}_\mu^{\ ij}{\rm L}_{ij}
\right) {\rm g}^{\mu\nu}|{\rm g}|^{1/2} \left(\partial_\nu-
\frac{i}{2}{\rm A}_\mu^{\ kl}{\rm L}_{kl} \right)+\\
\hskip1.2cm
+\frac{1}{2}\mathrm{II}_{i\mu\nu}\mathrm{II}^{i\mu\nu}-
\frac{1}{4}\mathrm{II}_{i\mu}^{\ \ \mu}\mathrm{II}_{\ \nu}^{i\
\nu}
 +\partial^i\partial_i+{\cal O}(\epsilon)
\end{array}\tag{\ref{LapRecuction}{\rm emb}}\label{Laplace emb}
\end{equation}
where ${\rm L}_{ij}=-i(y_i\partial_j-y_j\partial_i)$ are orbital
angular momentum operators in internal directions.}

\subsubsection{Dirac operator}
The HD Dirac operator $\bm{\mathcal D\!\!\!\!/}$ acts on
$2^{[D/2]}$-dimensional Dirac fermions. In every local coordinate
frame $\bm{\mathcal D\!\!\!\!/}$ is written in terms of HD gamma
matrices $\bm\gamma^A$, reference frames, partial derivatives,
pseudo-orthogonal connection coefficients and spin
pseudo-orthogonal generators
$\bm\Sigma^{AB}=-\frac{i}{4}[\bm\gamma^A,\bm\gamma^B]$ as
\begin{equation}
\bm{\mathcal D\!\!\!\!/}=\bm\gamma^C{\mathbf r}_C^{\ I}
\left(\partial_I-\frac{i}{2}{\mathbf\Omega}_{I,AB}\bm\Sigma^{AB}\right)
\nonumber
\end{equation}
$\bm{\mathcal D\!\!\!\!/}$ is Hermitian with respect to the
measure constructed by means of Dirac adjoint and HD covariant
volume element $|{\mathbf g}|^{1/2}d{\mathbf x}$. HD gamma
matrices $\bm\gamma^A$ can be decomposed in terms of LD external
$\gamma^\alpha$ and internal $\gamma^a$ gamma matrices as
\begin{eqnarray}
\bm{\gamma}^\alpha &=&\gamma^\alpha \otimes \bm{1}^{\rm int}\nonumber\\
\bm{\gamma}^a&=&\gamma^{\rm ext} \otimes \gamma^a\nonumber
\end{eqnarray}
where here and in what follows, $\bm{1}^{\rm ext}$, $\gamma^{\rm ext}$ and
$\bm{1}^{\rm int}$, $\gamma^{\rm int}$ denote identity and chiral matrices
in external and internal spin spaces, respectively. Correspondingly, the  HD
spin generators $\bm\Sigma^{AB}$ decompose in terms of LD external
$\Sigma^{\alpha\beta}=-\frac{i}{4}[\gamma^\alpha,\gamma^\beta]$ and internal
$\Sigma^{ab}=-\frac{i}{4}[\gamma^a,\gamma^b]$ ones as
\begin{eqnarray}
\bm\Sigma^{\alpha\beta} &=& \Sigma^{\alpha\beta}
\otimes {\mathbf 1}^{\rm int}\nonumber\\
\bm\Sigma^{\alpha b} &=& \frac{i}{2}
\gamma^{\rm ext}\gamma^\alpha \otimes\gamma^b\nonumber\\
\bm\Sigma^{ab} &=& {\mathbf 1}^{\rm ext} \otimes
\Sigma^{ab}\nonumber
\end{eqnarray}
By recalling the reference frames decomposition (\ref{rf}), the
relation between HD and LD connection coefficients
(\ref{HvLconnection}), suppressing --as customary-- tensor product
symbols and spin identity matrices, we obtain the most general LD
decomposition covariant under (\ref{STr})
\begin{eqnarray}
\bm{\mathcal D\!\!\!\!/}=\gamma^\gamma r_\gamma^{\ \mu}
\left(\hat\partial_\mu \!+\!\frac{1}{2}{E}_{\mu i}^{\ \
i}\!-\!\frac{i}{2} \hat{A}_{\mu,ab} \Sigma^{ab}\!-\!
\frac{i}{2}{{\mathit{\hat\Omega}}}_{\mu,\alpha\beta}
\Sigma^{\alpha\beta}\right) \!+
\nonumber\\
+\gamma^{\rm ext} \gamma^c \rho_c^{\ i}\left(\partial_i
\!+\!\frac{1}{2}\hat{E}_{i\mu}^{\ \
\mu}\!-\!\frac{i}{2}A_{i,\alpha\beta}\Sigma^{\alpha\beta}\!-\!\frac{i}{2}
{\mathit\Omega}_{i,ab}\Sigma^{ab}\right)\!+\nonumber\\
+\frac{i}{2}\gamma^{\rm ext}
\gamma^cf_{c\alpha\beta}\Sigma^{\alpha\beta}
\hskip0,5cm\label{DirRecuction}
\end{eqnarray}
The first righthand side term reproduces the four dimensional
Dirac operator
\begin{equation}
{\mathcal D\!\!\!\!/}^{\rm\ ext}= \gamma^\gamma r_\gamma^{\ \mu}
\left(\partial_\mu- \frac{i}{2}
{{\mathit{\Omega}}}_{\mu,\alpha\beta}
\Sigma^{\alpha\beta}\right)\nonumber
\end{equation}
with connection coefficients replaced by hatted ones and partial
derivatives $\partial_\mu$ replaced by
\[
\hat\partial_\mu +\frac{1}{2}{E}_{\mu i}^{\ \ i} - \frac{i}{2}
\hat{A}_{\mu,ab} \Sigma^{ab}
\]
As in the scalar case, the total derivative hidden in the trace of the
second fundamental form $\frac{1}{2}{E}_{\mu i}^{\ \ i}$ corrects the
different HD and LD normalization, while the operator gauge potential
$a_\mu-\frac{i}{2}\nabla_ia_\mu^i$ is now supplemented by the Hermitian
internal spin matrix  $\frac{1}{2}
\hat{A}_{\mu,ab} \Sigma^{ab}$.   The second righthand side term
corresponds to $\gamma^{\rm ext} $ times the internal Dirac
operator
\begin{equation}
{\mathcal D\!\!\!\!/}^{\rm\ int}= \gamma^c \rho_c^{\
i}\left(\partial_i -\frac{i}{2}
{\mathit\Omega}_{i,ab}\Sigma^{ab}\right) \nonumber
\end{equation}
with partial derivatives replaced by
\[
\partial_i+\frac{1}{2}\hat{E}_{i\mu}^{\ \
\mu}-\frac{i}{2}A_{i,\alpha\beta}\Sigma^{\alpha\beta}
\]
Once again $\frac{1}{2}\hat{E}_{i\mu}^{\ \ \mu}$ remedies the
different states normalization, while the Hermitian external spin
matrix $\frac{1}{2}A_{i,\alpha\beta}\Sigma^{\alpha\beta}$ enters
the expression as a gauge potential. The third righthand term
$\frac{i}{2}\gamma^{\rm ext}
\gamma^cf_{c\alpha\beta}\Sigma^{\alpha\beta}$ is an induced
Pauli term. Specializing to Kaluza-Klein and embedded spacetime
models we obtain:

 \vskip0,2cm\noindent{\small {\bf Kaluza-Klein} By
recalling (\ref{rfKK}), (\ref{IIexKK}), (\ref{LDexRKK}),
(\ref{IIinKK}), we have $\nabla_i{\rm K}_{\sf a}^i=0$ and assuming
that the external metric only depends on external coordinates,
(\ref{DirRecuction}) reduces to the well known Kaluza-Klein
decomposition of the Dirac operator
\begin{equation}
\begin{array}{l}
\bm{\mathcal D\!\!\!\!/}^{\rm KK}=\gamma^\alpha{\rm r}_\alpha^{\
\mu}\left(\partial_\mu - i{\rm A}_\mu^{\sf a}\hat{\rm K}_{\sf a}-
\frac{i}{2}\Omega_{\mu,\alpha\beta}\Sigma^{\alpha\beta}\right)+\\
\hskip0.4cm + \gamma^{\rm ext}\gamma^ak_a^{\ i}\left(\partial_i-
\frac{i}{2}\Omega_{i,ab}\Sigma^{ab}\right) +\frac{i}{2}\gamma^{\rm
ext}\gamma^i{\rm F}_{\alpha\beta}^{\sf a} {\rm K}_{{\sf
a}i}\Sigma^{\alpha\beta}
\end{array}
\tag{\ref{DirRecuction}KK}\label{Dirac KK}
\end{equation}
where
\begin{equation}
\hat{\rm K}_{\sf a}= -i{\rm K}^i_{\sf a}\partial_i+\frac{1}{2}
\left[k_a^{\ i}(\partial_i{\rm K}^j_{\sf a})k_{bj}-{\rm K}^i_{\sf
a}(\partial_i k_a^{\ j})k_{bj}\right]\Sigma^{ab}\nonumber
\end{equation}
are infinite dimensional Hermitian generators of the isometry
group algebra.}
\vskip0,05cm \noindent{\small {\bf Embedded spacetime} By
recalling (\ref{rfES}), (\ref{IIexES}), (\ref{LDexRES}),
(\ref{IIinES}), that $\nabla_i{\rm A}_j^{\ i}y^j=0$ and by
rescaling  fields and operators by
\begin{equation}
\begin{array}{rcl}
\bm\Psi & \rightarrow & |g|^{1/4}|{\rm g}|^{-1/4}\bm\Psi\\
\bm{\mathcal D\!\!\!\!/}\ & \rightarrow & |g|^{1/4}|{\rm
g}|^{-1/4} \bm{\mathcal D\!\!\!\!/}\ |g|^{-1/4}|{\rm g}|^{1/4}
\end{array}
\nonumber
\end{equation}
we obtain the following expression for the Dirac operator in
neighborhood of radius $\epsilon$ of ${\mathrm M}_d$
\begin{equation}
\begin{array}{l}
\bm{\mathcal D\!\!\!\!/}^{\rm\ emb}=\gamma^\alpha{\rm t}_\alpha^{\
\mu}\left(\partial_\mu- \frac{i}{2}{\rm A}_\mu^{\ ij}{\rm J}_{ij}-
\frac{i}{2}\Omega_{\mu,\alpha\beta}\Sigma^{\alpha\beta}\right)+\\
\hskip0.4cm+\gamma^{\rm ext}\gamma^i\partial_i +{\cal O}(\epsilon)
\end{array}
\tag{\ref{DirRecuction}{\scriptsize\rm emb}}\label{Dirac emb}
\end{equation}
with ${\rm J}_{ij}={\rm L}_{ij}+\Sigma_{ij}$ the total angular
momentum in internal directions.}
\subsubsection{Higher spin operators}
HD higher spin operators decompose in the very same way as the sum
of LD spin operators, with partial derivatives replaced by `gauge
covariant' ones and the possible addition of scalar potential
terms. In particular, external partial derivatives $\partial_\mu$
are replaced by
\begin{equation}
\hat\partial_\mu+\frac{1}{2}{E}_{\mu i}^{\ \ i} -\frac{i}{2}
\hat{A}_{\mu,ab}{\mathrm S}^{ab} \label{hs_gauge_derivative}
\end{equation}
with ${\mathrm S}^{ab}$ appropriate internal spin generators.

%%%%%%%%%%%%%%%%%%%%%%%%%%%%%%%%%%%%%%%%%%%%%%%%%%%%%%%%%%%%%%%
%
\section{\label{sec3} Gauge Symmetries from \protect\\
                      Higher Dimensional Covariance}
%
%%%%%%%%%%%%%%%%%%%%%%%%%%%%%%%%%%%%%%%%%%%%%%%%%%%%%%%%%%%%%%%%

One of the most interesting features of dimensional reduction is the
possibility of geometrically inducing gauge structures in the effective LD
dynamics. In this section we see that, after general covariance breaking,
residual internal coordinates and reference frame transformations are always
perceived by effective LD observers as gauge transformations. We also
discuss conditions for the induced gauge group to be finite dimensional,
providing a covariant characterization of Kaluza-Klein and other few
remarkable backgrounds.\\
Gauge fields are identified by their coupling to matter and by their
transformation rules. From the point of view of classical equation of
motion, a quick look to (\ref{H}) shows that $a_\mu^i\pi_i$ enters the
Hamiltonian as a gauge potential. To make the gauge structure explicit we
may rewrite the interaction term as $\mathbf{tr}(qa_\mu)$ with $q=iy^j\pi_j$
a suitable charge operator and $a_\mu=-ia_\mu^i\partial_i$ the gauge
connection introduced in Section \ref{sec1.2}. The corresponding curvature
$f_{\mu\nu}=-if_{\mu\nu}^i\partial_i$ enters the third term of equation
(\ref{geoex}). From the operatorial/quantum viewpoint, after adapting the
state measure to the external spacetime by the scale transformation
\begin{equation}
\begin{array}{l}
\bm\Psi\rightarrow|h|^{1/4}\bm\Psi\hskip0,3cm\mbox{and} \nonumber\\
{\mathbf\Delta}\rightarrow|h|^{1/4} {\mathbf\Delta}|h|^{-1/4},
\hskip0,2cm \bm{\mathcal D\!\!\!\!/}\rightarrow|h|^{1/4}
\bm{\mathcal D\!\!\!\!/}|h|^{-1/4} \hskip0,2cm  ...
\end{array}\nonumber
\end{equation}
expressions (\ref{LapRecuction}), (\ref{DirRecuction}) and
(\ref{hs_gauge_derivative}) taken by Laplace, Dirac and higher
spin operators show that
\begin{eqnarray}
{\mathcal A}_\mu&=&
-ia_\mu^i\left(\partial_i-\frac{i}{2}{\mathit\Omega}_{i,ab}
{\mathrm
S}^{ab} \right) -\frac{i}{2}\nabla_ia_\mu^i+\nonumber\\
& & +\frac{1}{2}(\partial_\mu\rho_a^{\ k})\rho_{bk}{\mathrm
S}^{ab} \label{gauge_potential}
\end{eqnarray}
couples to effective LD degrees of freedom as a gauge potential.
Under the residual covariance group  ${\mathcal A}_\mu$
transforms like  a gauge potential:
\begin{description}
\item  -- internal diffeomorphisms
\begin{equation}
y^i\rightarrow \exp\{\xi^k(x,y)\partial_k\}y^i \nonumber
\end{equation}
make $a_\mu^i$ and hence ${\mathcal A}_\mu$ to transform like
\begin{equation}
{\mathcal A}_\mu\rightarrow {T}{\mathcal A}_\mu
{T}^{-1}+i{T}(\partial_\mu {T}^{-1})
\end{equation}
with $T=\exp\{-\xi^k\partial_k\}$
\item  -- internal reference frame redefinitions
\begin{equation}
\rho_a^{\ i} \rightarrow \Lambda_a^{\ b}(x,y)\rho_b^{\ i}\nonumber
\end{equation}
make $(\partial_\mu\rho_a^{\ k})\rho_{bk}$ and hence ${\mathcal
A}_\mu$ to transform like
\begin{equation}
{\mathcal A}_\mu\rightarrow {\mathit \Lambda}{\mathcal A}_\mu
{\mathit \Lambda}^{-1}+i{\mathit \Lambda}(\partial_\mu {\mathit
\Lambda}^{-1})
\end{equation}
with ${\mathit \Lambda}=\exp\{\frac{i}{2} \Lambda_{ab}{\mathrm
S}^{ab}\}$
\end{description}
The commutator of two gauge covariant derivatives defines the operator
${\mathcal F}_{\mu\nu}=\partial_\mu {\mathcal A}_\nu-\partial_\nu{\mathcal
A}_\mu-i[{\mathcal A}_\mu,{\mathcal A}_\nu]$. A direct computation yields
\begin{eqnarray}
{\mathcal F}_{\mu\nu} &=& -i
f_{\mu\nu}^i\left(\partial_i-\frac{i}{2}{\mathit\Omega}_{i,ab}
{\mathrm S}^{ab} \right)
-\frac{i}{2}\nabla_if_{\mu\nu}^i+\nonumber\\
&&+\frac{1}{2}(\nabla_af_{b\mu\nu}){\mathrm S}^{ab}+{E}_{\mu a}^{\
\ c}{E}_{\nu bc} {\mathrm S}^{ab}\label{gauge_field}
\end{eqnarray}
Under internal diffeomorphisms and reference frames redefinitions
${\mathcal F}_{\mu\nu}$ correctly transforms like a gauge
curvature
\begin{equation}
{\mathcal F}_{\mu\nu}\rightarrow{T}{\mathcal F}_{\mu\nu}{T}^{-1}
\hskip0,3cm\mbox{and}\hskip0,3cm {\mathcal
F}_{\mu\nu}\rightarrow{\mathit \Lambda}{\mathcal
F}_{\mu\nu}{\mathit \Lambda}^{-1}
\end{equation}
${\mathcal A}_\mu$ and ${\mathcal F}_{\mu\nu}$ are Hermitian
operators --i.e.\ infinite dimensional matrices-- acting on
internal tensors/spinors. After HD covariance braking, residual
internal covariance is perceived by effective LD observers as an
infinite dimensional gauge group, with internal coordinate and
spin playing the role of --one of the many possible choices of--
gauge indices.
 The gauge curvature ${\mathcal F}_{\mu\nu}$ receives contributions
from two independent LD tensors: $f_{\mu\nu}^i$ and ${E}_{\mu ij}$. In
general, the two contributions are simultaneously active producing an
effective infinite-dimensional gauge group. In some special backgrounds the
gauge group may reduce to finite dimensions.

%%%%%%%%%%%%%%%%%%%%%%%%%%%%%%%%%%%%%%%%%%%%%%%%%%%%%%%%%%%%%%%%%%%
\subsection{\label{sec4.1}  ${E}_{\mu ij}=0$: gauge structures related \\
to the isometric structure of internal spaces}
%%%%%%%%%%%%%%%%%%%%%%%%%%%%%%%%%%%%%%%%%%%%%%%%%%%%%%%%%%%%%%%%%%%

Let us first consider the case where the internal fundamental form ${E}_{\mu
ij}$ vanishes identically while $f_{\mu\nu}^i$ is arbitrary. This
requirement is equivalent to the statement that the induced gauge structure
is of the Kaluza-Klein type. We have already seen in Section \ref{sec1.3.2},
equations (\ref{IIinKK}) and (\ref{IIinES}), that gauge structures of the
Kaluza-Klein type imply ${E}_{\mu ij}=0$. To prove the inverse, we note that
under the vanishing of the internal second fundamental form equation
(\ref{IIinIdentity}) implies that
\begin{equation}
\nabla_if_{j\mu\nu}+\nabla_jf_{i\mu\nu}=0\nonumber
\end{equation}
In every internal space the vector $f_{\mu\nu}^i$ is Killing. In principle
the Killing structure of  ${\mathrm M}_c^x$ can depend on the external point
$x$. However, the fact that $f_{\mu\nu}^i$ belongs to the Killing algebra
also implies that $a_\mu^i$ takes values on the same algebra up to a pure
gauge term. It is therefore possible to choose internal coordinates in which
$a_\mu^i$ is Killing, $\nabla_ia_{\mu j}+\nabla_ja_{\mu i}=0\nonumber$. In
such adapted coordinate frames equation (\ref{IIinKK'}) implies
$\partial_\mu h_{ij}=0$, that is $h_{ij}(x,y)=\kappa_{ij}(y)$. Thus, the
intrinsic geometry of internal spaces does not depend on the external
spacetime point. Having the same intrinsic and extrinsic geometry, all
internal spaces are isomorphic: ${\mathrm M}_c^x\equiv{\mathcal K}_c$. By
choosing a Killing vector basis ${\mathrm K}_{\mathsf a}^i(y)$, ${\mathsf
a}=1,...,n$, for the isometry algebra $iso({\mathcal K}_c)\equiv {\mathrm
g}_{\mathrm{KK}}$, $[{\mathrm K}_{\mathsf a},{\mathrm K}_{\mathsf
b}]^i=k_\mathsf{ab}^{ \ \mathsf{c}}{\mathrm K}_{\mathsf c}^i$, the
off-diagonal metric term $a_\mu^i$ and the antisymmetric hybrid tensor
$f_{\mu\nu}^i$ can be expanded as in (\ref{fKK}) or  (\ref{fES}) by
\begin{eqnarray}
a^i_\mu(x,y)&=&{\mathrm A}^{\mathsf a}_\mu(x){\mathrm K}_{\mathsf a}^i(y)\nonumber\\
f^i_{\mu\nu}(x,y)&=&{\mathrm F}^{\mathsf a}_{\mu\nu}(x){\mathrm
K}_{\mathsf a}^i(y)\nonumber
\end{eqnarray}
with ${\mathrm F}^{\mathsf c}_{\mu\nu}=\partial_\mu{\mathrm A}^{\mathsf
c}_\nu-\partial_\nu{\mathrm A}^{\mathsf c}_\mu-k_\mathsf{ab}^{ \
\mathsf{c}}{\mathrm A}^{\mathsf a}_\mu{\mathrm A}^{\mathsf b}_\nu$. The
gauge potential (\ref{gauge_potential}) and the gauge field
(\ref{gauge_field}) acting on  spin-$\mathrm{s}$ matter take then the
standard Kaluza-Klein form
\begin{eqnarray}
{\mathcal A}_{\mu}&=&{\mathrm A}_\mu^{\mathsf a}(x)\hat{\mathrm
K}_{\mathsf a}\label{KKpotential}\\ {\mathcal F}_{\mu\nu}&=&
{\mathrm F}^{\mathsf a}_{\mu\nu}(x)\hat{\mathrm K}_{\mathsf
a}\label{KKfiled}
\end{eqnarray}
with
\begin{eqnarray}
\hat{\mathrm K}_{\mathsf a}\!=\!{\mathrm K}_{\mathsf
a}^i\left(\!-i\partial_i\! -\!\frac{1}{2}\Omega_{i,ab}{\mathrm
S}^{ab}\!\right)\!+\!\frac{1}{2} (\nabla_a{\rm K}_{{\sf
a}b}){\mathrm S}^{ab}\nonumber
\end{eqnarray}
spin-$\mathrm{s}$ valued Hermitian differential operators closing the
finite-dimensional algebra $iso({\mathcal K}_c)$, $[\hat{\mathrm K}_{\mathsf
a},\hat{\mathrm K}_{\mathsf
b}]=-ik_\mathsf{ab}^{ \ \mathsf{c}}\hat{\mathrm K}_{\mathsf c}$.\\
The theory is still covariant under the whole residual covariance group
(\ref{STr}). However, a generic diffeomorphism
$T=\exp\{-\xi^i(x,y)\partial_i\}$ will bring $a_\mu^i$ outside
$iso({\mathcal K}_c)$. To keep the group structure of the background
explicit it is necessary to work in adapted coordinates.  This is achieved
by restricting the allowed covariance group to Killing transformations, that
is, by restricting attention to $\xi^i(x,y)=\epsilon^{\mathsf a}(x){\mathrm
K}_{\mathsf a}^i(y)$ as standard in Kaluza-Klein theories. In arbitrary
coordinate frames, Kaluza-Klein gauge structures are completely
characterized by the LD covariant condition
\begin{equation}
{E}_{\mu ij}=0
\end{equation}
Kaluza-Klein backgrounds, in the strict sense, further require the
independence of the induced external metric on internal coordinates, a
condition enforced by the vanishing of the symmetric part of the external
fundamental form $\hat{E}_{i(\mu\nu)}=0$. By contrast, with diffeomorphisms,
the Killing algebra of a manifold is always finite-dimensional having
dimension at most $c(c+1)/2$ \cite{Kill}. As a consequence, in the
Kaluza-Klein context, at least two internal dimensions are necessary to
produce non-Abelian gauge structures. Thus, a minimum of seven
extra-dimensions is required to realize the Standard Model group $U(1)\times
SU(2)\times SU(3)$
\cite{Witten81}.

%%%%%%%%%%%%%%%%%%%%%%%%%%%%%%%%%%%%%%%%%%%%%%%%%%%%%%%%%%%%%%%%%%%
\subsection{\label{sec4.2}  ${E}_{\mu ij}-\frac{1}{c}
{E}_{\mu k}^{\ \ k}h_{ij}=0$: gauge structures related \\
to the conformal structure of internal spaces}
%%%%%%%%%%%%%%%%%%%%%%%%%%%%%%%%%%%%%%%%%%%%%%%%%%%%%%%%%%%%%%%%%%%

Let us now weaken the Kaluza-Klein condition by requiring the
proportionality of the internal fundamental form ${E}_{\mu ij}$ to
the internal metric $h_{ij}$, ${E}_{\mu ij}=\frac{1}{c}{E}_{\mu
k}^{\ \ k}h_{ij}$ (this condition is trivial in $c=1$). Assuming
this, equation (\ref{IIinIdentity}) implies that
\begin{equation}
\nabla_if_{j\mu\nu}+\nabla_jf_{i\mu\nu}=\frac{2}{c}
(\nabla_kf^{k}_{\mu\nu})h_{ij}\nonumber
\end{equation}
In every internal space the internal vector $f^{i}_{\mu\nu}$ belongs to the
conformal algebra of the manifold. As above, the fact that $f^{i}_{\mu\nu}$
belongs to an algebra implies that also $a^i_\mu$ belongs to the same
algebra up to a pure gauge term. It is then possible to adapt internal
coordinates in such a way that $\nabla_ia_{\mu j}+\nabla_ja_{\mu
i}=\frac{2}{c}(\nabla_ka^k_\mu) h_{ij}$. Equation (\ref{IIinKK'}) implies
that $|h|^{1/c}\partial_\mu |h|^{-1/c}h_{ij}=0$. Hence, in the adapted
coordinate system $h_{ij}(x,y)=\lambda(x){\mathrm c}_{ij}(y)$ for some
conformal factor $\lambda(x)$ and some internal metric ${\mathrm
c}_{ij}(y)$. All internal spaces are conformal to a given manifold
${\mathcal C}_c$. Choosing a basis ${\mathrm C}_{\mathsf a}^i(y)$, ${\mathsf
a}=1,...,n$ for the conformal algebra $con\!f({\mathcal C}_c)$, $[{\mathrm
C}_{\mathsf a},{\mathrm C}_{\mathsf b}]^i=c_\mathsf{ab}^{ \
\mathsf{c}}{\mathrm C}_{\mathsf c}^i$, $a^i_\mu$ and
$f^{i}_{\mu\nu}$ can be expanded as
\begin{eqnarray}
a^i_\mu(x,y)&=&{\mathrm A}^{\mathsf a}_\mu(x){\mathrm C}_{\mathsf a}^i(y)\nonumber\\
f^i_{\mu\nu}(x,y)&=&{\mathrm F}^{\mathsf a}_{\mu\nu}(x){\mathrm
C}_{\mathsf a}^i(y)\nonumber
\end{eqnarray}
where again ${\mathrm F}^{\mathsf c}_{\mu\nu}=\partial_\mu{\mathrm
A}^{\mathsf c}_\nu-\partial_\nu{\mathrm A}^{\mathsf
c}_\mu-c_\mathsf{ab}^{ \ \mathsf{c}}{\mathrm A}^{\mathsf
a}_\mu{\mathrm A}^{\mathsf b}_\nu$. Also (\ref{gauge_potential})
and (\ref{gauge_field}) take the standard gauge potential and
gauge curvature form
\begin{eqnarray}
{\mathcal A}_{\mu}&=&{\mathrm A}_\mu^{\mathsf a}(x)\hat{\mathrm
C}_{\mathsf a}\label{CONFpotential}\\ {\mathcal F}_{\mu\nu}&=&
{\mathrm F}^{\mathsf a}_{\mu\nu}(x)\hat{\mathrm C}_{\mathsf
a}\label{CONFfiled}
\end{eqnarray}
where the spin-s valued Hermitian operators $\hat{\mathrm C}_{\mathsf a}$
take now the slightly more complicated form
\begin{equation}
\hat{\mathrm C}_{\mathsf a}\!= \!{\mathrm C}_{\mathsf
a}^i\left(\!-i\partial_i\!- \!\frac{1}{2}\Omega_{i,ab}{\mathrm
S}^{ab}\!\right)\!+ \!\frac{1}{2}(\nabla_a{\mathrm C}_{{\sf
a}b}){\mathrm S}^{ab}\!-\!\frac{i}{2}\nabla_i{\mathrm C}_{\sf a}^i
\nonumber%\label{CONFgen}
\end{equation}
It is readily checked that the $\hat{\mathrm C}_{\mathsf a}$ do not depend
on external coordinates and close $con\!f({\mathcal C}_c)$, $[\hat{\mathrm
C}_{\mathsf a},\hat{\mathrm C}_{\mathsf
b}]=-ic_\mathsf{ab}^{ \ \mathsf{c}}\hat{\mathrm C}_{\mathsf c}$.\\
As in the previous case, gauge invariance is only explicit when
the allowed covariance group is restricted to conformal
transformations $\xi^i(x,y)=\epsilon^{\mathsf a}(x){\mathrm
C}_{\mathsf a}^i(y)$, while in arbitrary coordinates the
background is completely characterized by the LD covariant
condition
\begin{equation}
{E}_{\mu ij}-\frac{1}{c}{E}_{\mu k}^{\ \ k}h_{ij}=0
\end{equation}
The conformal algebra of a manifold contains the isometry algebra as
subalgebra and is always finite dimensional with maximal dimension
$(c+1)(c+2)/2$. As a consequence non-Abelian gauge fields may be induced
even with a single internal dimension.
\vskip0,05cm \noindent{\small  {\bf Example:} To check this
explicitly we consider a one-dimensional internal space with topology of a
circle parameterized by the internal coordinate $\theta\in[-\pi,\pi]$. The
corresponding conformal algebra $so(2,1)$ is generated by the vector fields
${\mathrm C}_{\mathsf 1}^\theta=1$, ${\mathrm C}_{\mathsf
2}^\theta=\sin\theta$ and ${\mathrm C}_{\mathsf 3}^\theta=\cos\theta$.
Assuming the off-diagonal term of the HD metric to be of the form
\begin{equation}
a_\mu^\theta(x,\theta)=A_\mu^1(x)+A_\mu^2(x)\sin\theta+A_\mu^3(x)\cos\theta\nonumber
\end{equation}
the vector field (\ref{gauge_field}) rewrites like in
(\ref{CONFfiled}) with
\begin{eqnarray}
\hat{\mathrm C}_1&=&-i\frac{\partial}{\partial\theta}\nonumber\\
\hat{\mathrm C}_2&=&-i\sin\theta \frac{\partial}{\partial\theta}-
\frac{i}{2}\cos\theta\nonumber\\
\hat{\mathrm C}_3&=&-i\cos\theta \frac{\partial}{\partial\theta}+
\frac{i}{2}\sin\theta\nonumber
\end{eqnarray}
which are easily checked to close the $so(2,1)$ algebra
\begin{eqnarray}
[\hat{\mathrm C}_{\mathsf 1},\hat{\mathrm C}_{\mathsf
2}]=i\hat{\mathrm C}_{\mathsf 3},\hskip0,2cm [\hat{\mathrm
C}_{\mathsf 2},\hat{\mathrm C}_{\mathsf 3}]=-i\hat{\mathrm
C}_{\mathsf 1}, \hskip0,2cm [\hat{\mathrm C}_{\mathsf
3},\hat{\mathrm C}_{\mathsf 1}]=i\hat{\mathrm C}_{\mathsf 2}
\nonumber
\end{eqnarray}
We should remark, however, that $so(2,1)$ is the only non-Abelian
Lie algebra that can be embedded in $di\!f\!f({\mathrm M}_{1})$. }

%%%%%%%%%%%%%%%%%%%%%%%%%%%%%%%%%%%%%%%%%%%%%%%%%%%%%%%%%%%%%%%%%%%
\subsection{\label{sec4.3}  $f_{\mu\nu}^i=0$: gauge structures
related to the local\\ freedom of choosing internal reference
frames}
%%%%%%%%%%%%%%%%%%%%%%%%%%%%%%%%%%%%%%%%%%%%%%%%%%%%%%%%%%%%%%%%%%%

Eventually, we consider the case where the antisymmetric hybrid
tensor $f_{\mu\nu}^i$ vanishes identically while ${E}_{\mu ij}$ is
arbitrary. Under these circumstances it is always possible to
choose internal coordinates in such a way that the off-diagonal
block of the HD metric vanishes identically $a_\mu^i=0$. In such
adapted coordinate systems the internal fundamental form reduces
to the external derivative of the internal metric
\begin{equation}
{E}_{\mu ij}=\frac{1}{2}\partial_\mu h_{ij} \nonumber
\end{equation}
The gauge potential (\ref{gauge_potential}) and the gauge
curvature (\ref{gauge_field}) acting on  spin-$\mathrm{s}$ matter
take the form of standard (pseudo-)orthogonal gauge fields, with
internal spin generators ${\mathrm S}^{ab}$ playing the role of
gauge algebra generators
\begin{eqnarray}
{\mathcal A}_{\mu}&=&\frac{1}{2}(\partial_\mu\rho_a^{\
k})\rho_{bk}
{\mathrm S}^{ab}\\
{\mathcal F}_{\mu\nu}&=&\frac{1}{4}\rho_a^{\ i}\rho_b^{\
j}(\partial_\mu h_{ik})h^{kl}(\partial_\mu h_{jl})
% {E}_{\mu a}^{\ \ c}{E}_{\nu bc}
{\mathrm S}^{ab}
\end{eqnarray}
We see that LD gauge structures can be induced even when the
off-diagonal block of the HD metric vanishes identically, but they
only act on matter carrying spin. As in the previous cases, the
theory is still covariant under the whole residual covariance
group. The gauge structure emerges explicitly only when adapted
coordinates are introduced and the covariance group is restricted
to (pseudo-)rotations of internal reference frames. In generic
coordinate systems the background is fully characterized by the LD
covariant condition
\begin{equation}
f_{\mu\nu}^i=0
\end{equation}
Under these circumstances a minimum of three internal dimensions
is required to generate non-Abelian gauge structures, while ten
extra dimensions naturally provide the background for $SO(10)$
grand unification \cite{SO10}. Internal gauge indices like isospin
and color can  be nicely understood as internal spin indices  and
a complete matter unification can be achieved in terms of a single
fourteen-dimensional spinor \cite{Maraner04}.

%%%%%%%%%%%%%%%%%%%%%%%%%%%%%%%%%%%%%%%%%%%%%%%%%%%%%%%%%%%%%%%%%%
%
\section{\label{sec4} Discussion and Conclusion}
%
%%%%%%%%%%%%%%%%%%%%%%%%%%%%%%%%%%%%%%%%%%%%%%%%%%%%%%%%%%%%%%%%%%
The selection of a subset of coordinates --with the relative
general covariance breaking-- does not imply in itself neither the
selection of a reduced space nor a dimensional reduction
procedure. However, it determines the geometrical features of all
reduction schemes leading to that subset of coordinates as
residual coordinates.
 By investigating invariant/covariant quantities under the residual
transformation group we constructed LD tensors that fully characterize the
geometry of the coordinate choice and hence of the associated dimensional
reduction schemes. These allow to see in the same light reduction procedures
that seems otherwise totally unrelated, like Kaluza-Klein models --where the
system is totally delocalized in internal directions-- and embedded
spacetimes --where, on the contrary, the system gets localized at an
internal space point. Most of the formulas of Kaluza-Klein and embedded
spacetime theories do not depend on the averaging procedure employed, but
only on the geometry of the coordinate choice.
 In this paper we presented general formulas for the reduction of the
main tensors and operators of Riemannian geometry.
 In particular, the reduction of the HD Riemann tensor provides
what is probably the maximal possible generalization of Gauss,
Codazzi and Ricci equations.
 Our work also sheds some new light on the nature of geometrically
induced gauge structures, tracing their origin to residual
general covariance in internal directions.\\
\indent We conclude by remarking that --from the separation of
radial and angular coordinates in the two-body problem to the latest
theories of everything-- {\em adapting}, {\em selecting an appropriate
subset} and {\em exactly or effectively separating} coordinates is such a
basic procedure in solving physical problems, that it is unthinkable to
compile even a partial list of the papers where particular adapted/reduced
expressions of geometric tensors, equations and operators have been
obtained. Our hope is that the formulas presented in this paper may be of
help and save some tedious computational work to all researcher working on
some adapting coordinates problem.

\acknowledgements

J.K.P. would like to thank KITP for its hospitality.


\begin{thebibliography}{}
%
\bibitem{submanifolds} The geometry of embeddings is a
standard subject in differential geometry. Detailed discussions are found in
most textbooks. See for example L.~P.~Eisenhart,
 ``Riemannian Geometry'' (Princeton University Press, 1966);
B.~Y.~Chen,
 ``Geometry of Submanisolds'' (Dekker, 1973);
M.~Spivak,
 ``Differential Geometry'' Vol. 4 (Publish or Perish Inc., 1975);
%
\bibitem{embST} The idea that our four dimensional world can be an embedded object
in a higher dimensional spacetime, traces back at least as far as
D.~W.~Joseph,
 {\em Phys. Rev.} {\bf 126} (1962) 319 and {\em Rev. Mod. Phys.} {\bf 37} (1965) 225.
The subject was reinvented in the eighties by K.~Akama,
 in ``Lecture Notes in Physics'' Vol.\ 176 eds K.~Kikkawa,
 N.~Nakanishi and H.~Nariai (Springer Verlag, 1983);
V.~A.~Rubakov and M.~E.~Shaposhnikov,
 {\em Phys. Lett.} {\bf B125} (1983) 136;
M.~Visser,
 {\em Phys.\ Lett.\ } {\bf B159} (1985) 22;
G.~W.~Gibbson and D.~L.~Wiltshire,
 {\em Nuc. Phys.} {\bf B287} (1987) 717.
Recently the idea has received renewed attention in connection with the
induced-matter theory of P.~S.~Wesson,
 ``Space-Time-Matter'' (World Scientific, 1999)
and with the brane world scenario of N.~Arkani-Hamed, S.~Dimopoulos and
G.~Dvali,
 {\em Phys.\ Lett.\ } {\bf B429} (1998) and {\em Phys.\ Rev.\ } {\bf D59} (1999) 86004;
I.~Antoniadis, N.~Arkani-Hamed, S.~Dimopoulos and G.~Dvali,
 {\em Phys.\ Lett.\ } {\bf B436} (1998) 257;
L.~Randall and R.~Sundrum,
 {\em Phys.\ Rev.\ Lett.\ } {\bf 83} (1999) 3370 and {\em Phys. Rev. Lett.} {\bf 83} (1999) 4690.
See also Josep M. Pons and Pere Talavera, {\em Nucl. \ Phys.} {\bf B703}
(2004) 537;  C. Kokorelis,{\em Nucl. Phys.} {\bf B677} (2004) 115; D.
Cremades, L.E.Ibanez and F.Marchesano, {\em Nucl. Phys.} {\bf B643} (2002)
93.
%
\bibitem{KK}
T.~Kaluza,
 {\em Sitzungsber.\ d.\ Berl.\ Akad.\ } (1921) 996;
O.~Klein,
 {Z.\ F.\ Physik} {\bf 37} (1926) 895 and {\em Nature} {\bf 118} (1926) 516.
%
\bibitem{nonAbKK} The non-Abelian generalization of Kaluza-Klein
theory was addressed by B.~S.~De Witt
 in ``Relativity, Groups and Topology'', eds C. and B.~S.~De Witt
 (Gordan and Breach, New York, 1964)
followed by the work of J.~Rayski,
 {\em Acta.\ Phys.\ Polon.\ } {\bf 27} (1965) 947 and {\bf 28} (1965) 87;
R.~Kerner,
 {\em Ann.\ Inst.\ Poincar\'e}, Sect.\ {\bf A9} (1968) 29;
A.~Trautman,
 {\em Rep.\ Math.\ Phys.\ } {\bf 1} (1970) 29;
Y.~M.~Cho,
 {\em J.\ Math.\ Phys.\ } {\bf 16} (1975) 2029;
Y.~M.~Cho and P.~G.~O.~Freund,
 {\em Phys.\ Rev.\ } {\bf D12} (1975) 1711;
Y.~M.~Cho and P.~S.~Jang,
 {\em Phys.\ Rev.\ } {\bf D12} (1975) 3789
and others. Eventually J.~F.~Luciani,
 {\em Nucl.\ Phys.\ } {\bf B135} (1978) 111
realized that any internal space with isometry group ${\mathrm G}$, will
produce ${\mathrm G}$ as effective four dimensional gauge symmetry.
%
\bibitem{KKrev} Most of relevant historical papers on Kaluza-Klein theory are
reprinted in ``Modern Kaluza-Klein theories'', eds T.~Appelquist, A.~Chodos
and P.~G.~O.~Freund (Addison-Wesley, Menlo Park, 1987). Recent review on the
subject are found in D.~Bailin and A.~Love,
 {\em Rep. Prog. Phys.} {\bf 50} (1987) 1087;
M.~J.~Duff,
 {\tt hep-th/9410046} (1994);
J.~M.~Overdium and P.~S.~Wesson,
 {\em Phys. Rept.} {\bf 283} (1997) 303.
%
\bibitem{GCR} Generalization and applications of Gauss, Codazzi and Ricci equations were recently
discussed by R.~Capovilla and J.~Guven,
 {\em Phys.\ Rev.\ } {\bf D51} (1995) 6736
and
 {\em Phys.\ Rev.\ } {\bf D52} (1995) 1072;
B.~Carter,
 {\em Contemp.\ Math.\ } {\bf 203} (1997) 207;
B.~G.~Konopelchenko
 {\em J.\ Phys.\ A: Math.\ Gen.\ }{\bf 30} (1997) L437;
E.~Zafiris,
  {\em  Phys.\ Rev.\ }{\bf D58} (1998) 043509;
M.~D.~Maia and E.~M.~Monte,
 {\em J.\ Math.\ Phys.\ } {\bf 37} (1996) 1972
and
 {\em Phys. Lett.} {\bf A297} (2002) 9;
E.~Gourgoulhon and J.~L.~Jaramillo,
 {\em Phys.\ Rept.\ }{\bf 423}, 159 (2006);
S.~Pal and S.~Kar,
  {\em Class.\ Quant.\ Grav.\ }{\bf 23}, 2571 (2006);
S.~Kar and S.~SenGupta,
  {\tt arXiv:gr-qc/0611123}.
%
\bibitem{Weinberg72}
S.~Weinberg,
 ``Gravitation and Cosmology'' (John Wiley and Sons Inc., 1972).
%
\bibitem{gaugeEST}
The fact that the normal connection induced on a submanifold appears as a
gauge potential in effective brane world dynamics was first pointed out by
K.~Akama,
 {\em Prog.\ Theo.\ Phys.\ } {\bf 78} (1987) 184.
The idea was then independently rediscovered and/or discussed in various
contexts by P.~Maraner and C.~Destri,
 {\em Mod.\ Phys.\ Lett.\ } {\bf A8} (1993) 861;
K.~Fujii and N.~Ogawa,
 {\em Prog.\ Theo.\ Phys.\ } {\bf 89} (1993) 575;
P.~Maraner,
 {\em J. Phys. A: Math Gen.} {\bf 28} (1995) 2939
and
 {\em Ann.\ Phys.\ } {\bf 246} (1996) 325;
N.~M.~Chepilko and K.~Fujii,
 {\em Phys.\ Atom.\ Nuc.\ } {\bf 58} (1995) 1063;
K.~Fujii, N.~Ogawa, S.~Uchiyama and N.~M.~ Chepilko,
 {\em Int.\ J.\ Mod.\ Phys.\ } {\bf A12} (1997) 5235;
H.~Grundling and C.~A.~Hurst,
 {\em J.\ Math.\ Phys.\ } {\bf 39} (1998) 3091;
N.~Ogawa,
 {\tt hep-th/9801115};
K.~Akama,
 {\em Mod.\ Phys.\ Lett.\ } {\bf A15} (2000) 2017;
P.~C.~Schuster and R.~L.~Jaffe,
 {\em Ann.\ Phys.\ } {\bf 307} (2003) 132.
%
\bibitem{UtiKib}
R.~Utiyama,
 {\em Phys.\ Rev.\ } {\bf 101} (1956) 1597;
T.~W.~B.~Kibble,
 {\em J.\ Math.\ Phys.\ } {\bf 2} (1961) 212.
%
\bibitem{STM}
P.~S.~Wesson,
 {\em Gen.\ Rel.\ Grav.\ } {\bf 22} (1990) 707;
P.~S.~Wesson and J.~Ponce~de~Leon,
 {\em J.\ Math.\ Phys.\ } {\bf 33} (1992) 3883;
J.~Ponce~de~Leon and P.~S.~Wesson,
 {\em J.\ Math.\ Phys.\ } {\bf 34} (1993) 4080;
P.~S.~Wesson, J.~Ponce~de~Leon, P.~Lim and H.~Liu,
 {\em Int.\ J.\ Mod.\ Phys.\ } {\bf D2} (1993) 163;
P.~S.~Wesson,
 {\em Mod.\ Phys.\ Lett.\ } {\bf A10} (1995), 15;
S.~Rippl, C.~Romero and R.~Tavakol,
 {\em Class.\ Quant.\ Grav.\ } {\bf 12} (1995) 2411.
%
\bibitem{Jackiw06}
R.~Jackiw,
 in the Proceedings of the Seventh International Conference
 "Symmetry in Nonlinear Mathematical Physics" (Kyiv, Ukraine,
 2007),
 {\tt http:/www.emis.de/journals/SIGMA/2007/091/};
D.~Grumiller and R.~Jackiw,
 {\em Int.\ J.\ Mod.\ Phys.\ } {\bf D15} (2006) 2075.
 %
\bibitem{Kill} See for example
B.~O'Neill
 ``Semi-Riemannian Geometry'' (Academic Press, 1983)
%
\bibitem{Witten81}
E.~Witten,
 {\em Nuc.\ Phys.\ } {\bf B186} (1981) 412.
%
\bibitem{SO10}
H.~Geoergi,
 in {\em Particles and Fields}, ed. C. E.Caelson (AIP, 1975), p. 575;
%
\bibitem{Maraner04}
P.~Maraner,
 {\em Mod.\ Phys.\ Lett.\ } {\bf A19} (2004), 357.
%
\end{thebibliography}
\end{document}